\documentclass[12pt,reqno,a4paper]{amsart}

\usepackage[margin=3cm]{geometry}

\usepackage{amsmath}
\usepackage{amsfonts}
\usepackage{amssymb}
\usepackage{amscd}
\usepackage{amsthm}
\usepackage{enumerate}
\usepackage{graphicx}

     \newcommand{\tr}{{\operatorname{tr}}}

     \newcommand{\N}{{\mathbb{N}}}
     \newcommand{\R}{{\mathbb{R}}}
     
     \newcommand{\C}{{\mathbb{C}}}

\newcommand{\w}{{\rm w}}
\newcommand{\s}{{\rm s}}
\newcommand{\e}{{\rm e}}
\renewcommand{\i}{{\rm i}}
\renewcommand{\d}{{\rm d}}

\newcommand{\unif}{{\rm unif}}

\renewcommand{\tr}{{\rm tr}}

\renewcommand{\Re}{{\rm Re}\,}

     \theoremstyle{plain}
     \newtheorem{thm}{Theorem}[section]
     \newtheorem{prop}[thm]{Proposition}
     \newtheorem{lemma}[thm]{Lemma}
      \newtheorem{cor}[thm]{Corollary}
     \theoremstyle{definition}
     
     \newtheorem{defn}[thm]{Definition}
     \newtheorem{example}[thm]{Example}
     
     \newtheorem{cond}[thm]{Condition}

     \newtheorem{remark}[thm]{Remark}
     \newtheorem{remarks}[thm]{Remarks}
\newtheorem*{remarks*}{Remarks}
\newtheorem*{remark*}{Remark}

     \numberwithin{equation}{section}
\title{Classical scattering at low energies}

\author{J. Derezi\'nski}
\address[J. Derezi\'nski]{Dept. of Math. Methods in Physics, Warsaw University\\ 
Ho\.za 74, 00-682, Warszawa, Poland} 
\email{Jan.Derezinski@fuw.edu.pl}

\author{E. Skibsted}
\address[E. Skibsted]{Institut for  Matematiske
Fag \\
Aarhus Universitet\\ Ny Munkegade  8000 Aarhus C, 
Denmark}
\email{skibsted@imf.au.dk}
\thanks{The authors are  partially supported by  MaPhySto -- A
Network in Mathematical Physics and Stochastics, funded by
The Danish National Research Foundation
and  by  the Postdoctoral Training Program 
HPRN-CT-2002-0277. The research of J.D.
 was also supported by the Polish grants
 SPUB127 and  2 P03A 027 25. Part of the research was done during a visit of
 both authors to the Erwin
 Schr\"odinger Institute.
}

\begin{document}

\begin{abstract} For a class of negative  slowly decaying
  potentials  including the attractive Coulombic one we study the
  classical scattering theory in the low-energy regime. We construct  a (continuous) family of classical orbits parametrized by initial position $x\in \R^d$, final direction
  $\omega\in S^{d-1}$ of escape (to infinity) and the
  energy $\lambda\geq 0$, yielding  a complete classification
  of the set of outgoing scattering orbits. The construction is given in the
  outgoing part of phase-space (a similar construction may be done in
  the incoming part of phase-space). For fixed $\omega\in S^{d-1}$
  and $\lambda\geq 0$ the collection of constructed orbits constitutes
  a smooth manifold that we show is Lagrangian. The family of those
  Lagrangians can be used to study the quantum mechanical scattering
  theory in the low-energy regime for the class of potentials
  considered here. We devote this study to a subsequent paper \cite{DS}.
\end{abstract}

\maketitle

\tableofcontents
\section{Introduction} \label{Introduction}

In this paper we  shall address  a classical low-energy scattering problems
for a two-body system. In a subsequent paper \cite{DS} we shall
combine the results of this paper and some of \cite{FS} in  a 
study of the quantum mechanical low-energy scattering theory within the
same class of potentials; this will include construction of  wave operators, 
corresponding generalized eigenfunctions and $S$--matrices and the
establishment of 
 regularity/asymptotics of these quantities at the threshold energy
 $\lambda= 0$. A related programme has  been carried out for
 positive energies for a
 wide class of potentials, see \cite{IK1}, \cite{IK2},
\cite{Y2} and  \cite{RY}; for this case the study is based on a relatively simple
structure of a class of outgoing (or incoming) classical orbits used
in  the construction of certain Fourier integral operators. 

 However there are severe difficulties if one tries to incorporate the low energy regime
$\lambda\approx 0$ by 
this approach. Some of the difficulties already show up at the
classical level,  and therefore one needs additional  conditions on
the potential from the very outset of the analysis. In our
opinion  these ``additional  conditions''  naturally count the virial
condition and spherical  symmetry of the
leading term of the potential (both conditions to be imposed at
infinity only). 

To simplify the presentation let us in this introduction assume that the potential takes
the form (with $x\in \R^d$ for $d\geq 2$)
\begin{equation}
  \label{eq:1v}
  V(x)=-\gamma |x|^{-\mu} +O(|x|^{-\mu-\epsilon}),
\end{equation} where $\mu \in (0,2)$ and $\gamma, \epsilon>0$. For 
derivatives, assume that $\partial^\beta
\{V(x)+\gamma |x|^{-\mu}\}=O\big (|x|^{-\mu-\epsilon-|\beta|}\big ).$ 
 We look at
the  classical Hamiltonian 
$h(x,\xi)=\tfrac {1}{2 }\xi ^2+V(x)$.

With  \eqref{eq:1v} one can prove the existence of
the {\it asymptotic normalized
  velocities} for any {\it classical scattering orbit}, i.e. a solution to
Newton's equation with $|x(t)| \to
\infty$, 
\begin{equation}
  \label{eq:asymptotic norm222}
 \omega^{\pm}={\pm}\lim_{t\to {\pm}\infty}x(t)/|x(t)|;
\end{equation}  notice that this includes orbits  
with arbitrary energy $\lambda\geq 0$. In particular we see that there
is a  well-defined classical
scattering theory for $\lambda=0$ (the quantities $\omega^{\pm}$ are
outgoing and incoming directions).

We look at the following mixed problem (restricted to  outgoing and
incoming  regions): Consider the momentum 
$\xi=\sqrt{2\lambda}\omega$ as depending on the two independent
variables $\lambda\geq 0$ and $\omega\in S^{d-1}$ and solve 
\begin{equation}\label{eq:mixed conditions222}
\begin{cases}
\ddot y(t) =-\nabla V(y(t))\\
\lambda=\tfrac {1}{2 }\dot y(t)^2 +V(y(t))\\
y({\pm}1)=x\\
\omega={\pm}\lim_{t\to {\pm}\infty}y(t)/|y(t)|
\end{cases}\;.
\end{equation}

The bulk of the paper is devoted to solving \eqref{eq:mixed
  conditions222} and deriving various regularity properties. Given  solutions to \eqref{eq:mixed
  conditions222} we then construct phases 
$\phi^{\pm}(x,\omega,\lambda)$  in terms of  the velocity fields $$\nabla_x \phi^{\pm}(x,\omega,\lambda)=\dot
y({\pm}1)=\dot
y(t={\pm}1;x,\omega,\lambda).$$

It turns out that the constructed phases 
$\phi^{\pm}(x,\omega,\lambda)$  are jointly continuous but lack smoothness in
the $\lambda$--variable at $\lambda=0$. 

We give a complete classification of the set of scattering orbits: For any scattering orbit $x(t)$ with asymptotic
velocities $\omega^{\pm}$ given by \eqref{eq:asymptotic norm222} and
energy  $\lambda\geq 0$ there exists a
(large) $T_0>0$ such that for all $\pm t\geq T\geq T_0$
\begin{align*}
 x(t)&=y(t\mp T\pm 1; x(\pm T),\omega ^{\pm},\lambda),\\
\dot
x(t)&=\nabla_x \phi^{\pm}(x(t),\omega^{\pm},\lambda).
\end{align*}

A typical orbit $x(t)$ for $\lambda=0$ is depicted
 below, see Example \ref{example:hom-poten}
for an elaboration.

\vspace{0.5cm}
\begin{center}
\includegraphics[width=9cm,totalheight=7cm]{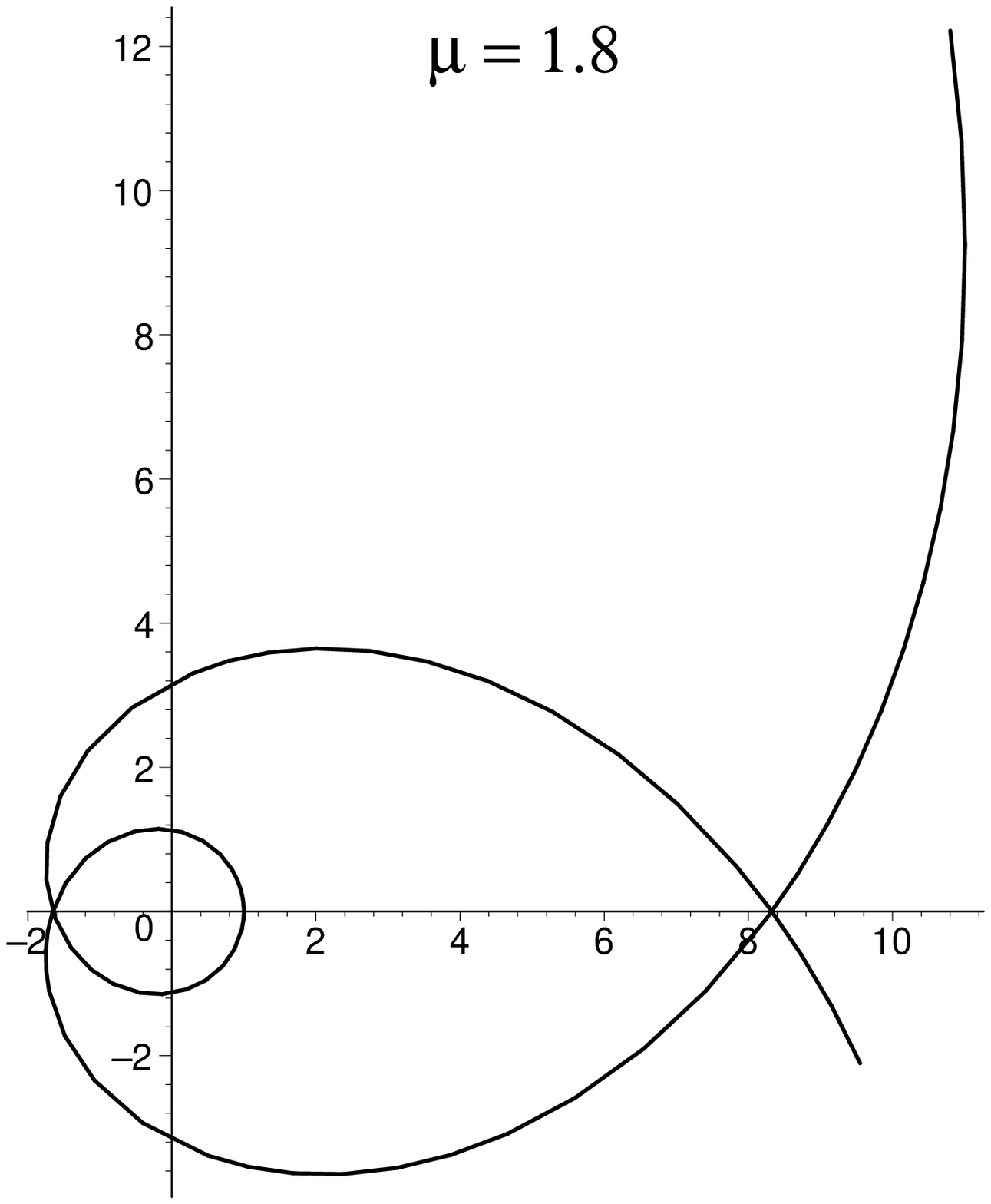}
\end{center}

The paper is organized as follows: In Section \ref{Conditions} we
impose conditions on the potential. In the case we allow the potential 
 to have  a
non-spherically symmetric term  we shall need  certain regularity properties
of the leading spherically symmetric term. These properties are stated
in Condition \ref{assump:conditions2}; they are fulfilled for the
example \eqref{eq:1v} discussed above.

In Section \ref{Asymptotic
  normalized velocity} we show the
existence of the asymptotic normalized
  velocity in the classical theory (only the $+\infty$ case is
  treated). 

In Sections
  \ref{Mixed problem in the case V_2= V_3= 0} and \ref{Mixed problem
    in the case V_2 not 0} we solve the   mixed problem  \eqref{eq:mixed
    conditions222} (the $+\infty$ case only), first in the  case of spherical symmetry and
  then in the more general non-spherically symmetric case, and we
  derive smoothness properties of the solution. The first case is
  treated by the implicit function theorem while our study of the
  second case is based on a perturbation and Taylor expansion argument 
  (similar to \cite {S1, S2}) allowing us to set
  up  a fixed point problem. The material is technically somewhat
  complicated, and to  improve the presentation we devote  Section
  \ref{Time--dependent linear force problem } to some (abstract)
  preliminaries for Section \ref{Mixed problem
    in the case V_2 not 0} related to the uncertainty principle lemma
  (Hardy inequality). The basic issue of  Section
  \ref{Time--dependent linear force problem } is a limiting absorption
  principle at zero energy for a one-dimensional vector-valued problem in which the
  time variable plays the role of a configuration space variable!

We
  prove in Section \ref{Solution to the eikonal equation} that the
  outgoing 
  velocity field $(x,F(x))$ from Definition \ref{def:vector field} is
  Lagrangian,
so that $F=\nabla \phi$ for some phase function $\phi$. 
Then we fix $\phi^+=\phi$  by specifying
its  value at a
(local) point. We also explain how to define $\phi^-$. These
  constructions  will be  the outset for studying   quantum mechanics in
  \cite {DS}. Finally we show that the family of orbits
  \eqref{eq:mixed conditions222} yields a complete classification of
  the set of scattering orbits.

\section{Conditions} 
\label{Conditions}
We shall consider a classical Hamiltonian $h=\tfrac {1}{2 }\xi^2+ V$ on ${\mathbb
R}^d\times {\mathbb R}^d$ where $d\geq 2$ and $V$ satisfies Condition
\ref{assump:conditions1} and possibly Conditions
\ref{assump:conditions2} and \ref{assump:conditions3} 
(all stated below). We shall use the
standard notation $\langle x \rangle=(1+x^2)^{1/2}$ for $x\in \R^d$.
\begin{cond}
\label{assump:conditions1}
The function $V$ can be written as a sum of two 
real-valued smooth  functions, $V= V_1 + V_2$, such that:
For some $\mu \in (0,2)$ we
have

\begin{enumerate}[\quad\normalfont (1)]

   \item \label{it:assumption1} $V_1$ is a negative function that only 
     depends on  the 
     radial variable $r=|x|$ in the region $r\geq 1$ (that is
     $V_1(x)=V_1(r)$ for $r\geq 1$). There exists $\epsilon_1 > 0$ such
     that $$V_1(r) \leq -\epsilon_1
r^{-\mu};\;r\geq 1.$$
   \item \label{it:assumption2}     For all $\gamma\in ({\mathbb N} \cup \{0\})^{d}$
there exists
$C_{\gamma} >0$ such that
$$
\langle x \rangle^{\mu+|\gamma|} |\partial^{\gamma} V_1(x)|
\leq C_{\gamma}.$$
\item \label{it:assumption3} 
     There exists $\tilde\epsilon_1 > 0$ such that 
     \begin{equation}
       \label{eq:virial}
       rV_1'(r)
\leq -(2-\tilde\epsilon_1) V_1(r);\;r\geq 1.
     \end{equation}

\item \label{it:assumption4} 
     There exists $\epsilon_2 > 0$ such that for all $\gamma\in ({\mathbb N} \cup \{0\})^{d}$
$$
\langle x \rangle^{\mu+\epsilon_2 +|\gamma|} |\partial^{\gamma} V_2(x)|
\leq C_{\gamma}.$$
\end{enumerate}
\end{cond}


We introduce the quantity
\begin{equation}
  \label{eq:tildtt}
  \tilde t (r)=\int^r_1 (-2V_1(\rho))^{-\frac
  {1}{2}}\,\d\rho;\;r\geq 1,
\end{equation} which is the time of arrival  at  distance $r$
  from  the origin  for a purely outgoing zero-energy
  orbit starting at $r=1$ at time $t=0$ (assuming $V_2=0$).

 The following condition will be  needed only in the case $V_2\neq
 0$. We notice that \eqref{eq:virial} and \eqref{eq:green's1} tend to
 be somewhat strong conditions for $\mu{\approx}2$. On the other hand  Conditions
 \ref{assump:conditions1} and \ref{assump:conditions2} hold for all
 $\epsilon_2>0$ for the particular example
 $V_1(r)=-\gamma r^{-\mu}$ (with $\epsilon_1 =\gamma$, $\tilde
 \epsilon_1=2-\mu$  and some $\bar\epsilon_1<1-\alpha
\mu$), cf. Section \ref{Introduction}.
\begin{cond}
\label{assump:conditions2}
Let $V_1$ be given as in Condition  \ref
{assump:conditions1}, and define $\alpha
=\tfrac{2}{2+\mu}$. There exists $\bar\epsilon_1 > \max (0,
1-\alpha(\mu+2\epsilon_2))$ such that 
\begin{align}
 \label {eq:green's1}
  \limsup_{r\to \infty} r^{-1}V_1'(r) \,\tilde t (r)^2 &< 4^{-1}(1-\bar\epsilon_1^2),\\
\limsup_{r\to \infty} V_1''(r)\,\tilde t (r)^2 &< 4^{-1}(1-\bar\epsilon_1^2).\label {eq:green's2}
\end{align}
\end{cond}

Let us for convenience assume under Condition
\ref{assump:conditions1} that 
\begin{equation}
  \label{eq:e_2small}
  \epsilon_2\leq 4^{-1}(2-\mu).
\end{equation} Notice that under Condition
\ref{assump:conditions2}, \eqref{eq:e_2small}  is not in conflict with the condition $\bar\epsilon_1 > \max (0,
1-\alpha(\mu+2\epsilon_2))$.
 
The following condition will be  needed only in 
Subsection \ref{Classification of scattering orbits}.

\begin{cond}
\label{assump:conditions3}
Let $V_1$, $V_2$ and $\bar \epsilon_1$ be given as in Conditions
\ref{assump:conditions1} and \ref{assump:conditions2}. Then
\begin{equation}
 \label {eq:green's3}
  \limsup_{r\to \infty} {-V_1'(r)\over \sqrt{-2V_1(r)}}\,\tilde t (r)
  < 2^{-1}(1+\bar\epsilon_1).
\end{equation}
\end{cond}

In the (typical) situation  where $V_1(r)$ is concave at infinity obviously the bounds
\eqref{eq:green's2} and \eqref{eq:green's3} are fulfilled.

We shall often use the
 notation $x=r\hat x$ with  $r=|x|$ and $\hat x=x/r$ for vectors $x\in
 \R^d\setminus \{0\}$.  
The notation $F(s>\epsilon)$ denotes 
a smooth increasing function $=1$ for $s>\frac {3}{4}\epsilon$ and
 $=0$ for $s<\frac {1}{2}\epsilon$;
 $F(\cdot<\epsilon):=1-F(\cdot>\epsilon)$.
Throughout the paper the notation $\mu$ refers to the number $\mu$
appearing in Condition \ref{assump:conditions1} and
$\alpha:=2/(2+\mu)$, cf. Condition 
\ref{assump:conditions2}. 
 The function  $g(r):=\sqrt{2\lambda-2V_1(r)}$
 (for $V_1$ obeying Condition \ref{assump:conditions1}) will also be
 used extensively. This quantity represents the speed of any orbit with
  energy $\lambda$ and located at  distance $r$
  from  the origin (assuming $V_2=0$).

\section{Asymptotic normalized velocity}
\label{Asymptotic normalized velocity}

     \newcommand{\crit}{{\rm tp}}     \newcommand{\al}{{\rm al}}

 We
define
a {\it classical outgoing scattering orbit}  to be a solution to 
Newton's  equation $\ddot x(t)=-\nabla V(x(t))$ obeying $|x(t)|\to
\infty$ for $t\to +\infty$ (we consider for convenience here only the case of $t\to
+\infty$).
In this section we investigate various general
 properties of scattering orbits.
Note that all the conditions
on the potentials used in this section are implied by
 Condition \ref{assump:conditions1}.

 The energy of the orbit is given by 
  \begin{equation*}
    \lambda=\tfrac {1}{2 }\dot x(t)^2 +V(x(t)).
\end{equation*}

We start with a well known consequence of the positivity of the virial.

\begin{prop} Suppose that for  $|x|\geq R$
\[-2V(x)-x\cdot \nabla V(x)\ge0.\]
Then for any outgoing scattering orbit there exists $T$ such that for $t\geq T$
\[x(t)\cdot \dot x(t)\geq 2(t-T)\lambda,\ \ \ 
x^2(t)\geq 2\lambda(t-T)^2+R^2.\]
\label{prop1}\end{prop}

\begin{proof}
We have
for $|x(t)|\geq R$
\begin{eqnarray}
\frac12
\frac{\d^2}{\d t^2}x^2(t)
 =
\frac{\d}{\d t}\big (x(t)\cdot \dot x(t)\big )&=&
2\lambda-2V(x(t))-x(t)\cdot \nabla V(x(t))\nonumber\\
&\geq& 2\lambda.\label{boo}
\end{eqnarray} 
If  $x(t)$ is a scattering orbit,  we can find $T$ such that
$\frac{\d}{\d t}x^2(T)\geq 0$ and $|x^2(T)|> R^2$.
So (\ref{boo}) is satisfied for all $t\geq T$, and the result follows
from integration. \end{proof}

The
following proposition can be traced back to \cite{Ge}, see also \cite[Theorem 4.7] {FS}.

\begin{prop} Suppose that 
\begin{equation}
2V(x)+x\cdot \nabla V(x)\leq -c|x|^{-\mu},\ \ \ c>0,\ \ \   |x|\geq R
\label{asso}\end{equation}
Then for any outgoing
 scattering orbit for large enough time and some $\epsilon>0$,
\begin{equation}\label{eq:alpha}
 |x(t)|\geq \epsilon t^\alpha. 
\end{equation}
\end{prop}

\begin{proof} 
For large enough $T$ and $t\geq T$
 we have $|x(t)|\geq R$. Then
\begin{eqnarray}
\frac12
\frac{\d^2}{\d t^2}x^2(t)
&=
2\lambda-2V(x(t))-x(t)\cdot \nabla V(x(t))
&\geq c|x(t)|^{-\mu}
.\label{boo1}
\end{eqnarray}
We multiply (\ref{boo1}) from both sides by $\frac{\d}{\d t }x^2(t)$ and, using $\mu<2$, we 
obtain
\[\frac{\d}{\d t}\left(\frac{\d}{\d t}x^2(t)\right)^2\geq
c_1\frac{\d}{\d t} \left(x^2(t)\right)^{1-\frac{\mu}{2}}.\]
This yields
\[\left(\frac{\d}{\d t}x^2(t)\right)^2\geq
c_1 \left(x^2(t)\right)^{1-\frac{\mu}{2}}+c_2.\]
By Proposition \ref{prop1} we know that for large  times 
$\frac{\d}{\d t}x^2(t)\geq0$ is positive, and
hence
\[\frac{\d}{\d t}x^2(t)\geq
\left(c_1 \left(x^2(t)\right)^{1-\frac{\mu}{2}}+c_2\right)^{\frac12}.\]

This implies  for large enough time (\ref{eq:alpha}).
\end{proof}

The  upper 
bound on the zero energy orbit (\ref{eq:examlog00}) 
can be traced back to
\cite{De1,De2}. 
\begin{prop} Assume that $ V(x)=O(|x|^{-\mu})$. Then
 the outgoing scattering orbits with $\lambda=0$ satisfy the bound
\begin{equation}\label{eq:examlog00}
  x(t)=O(t^{\alpha}).
\end{equation} 
If in addition,  for $|x|\geq R$, $V(x)\leq -c_0 |x|^{-\mu}$, $c_0>0$,
then all outgoing scattering 
orbits for large enough time satisfy the bound
\begin{equation}
\label{eq:velocity bound}
 |\dot x(t)|\geq \epsilon t^{\alpha-1}. 
\end{equation}
If also for (\ref{asso}) holds, then
 the  orbits with $\lambda=0$  satisfy
\begin{equation}\label{eq:examlog000}
  \dot x(t)=O(t^{\alpha-1}).
\end{equation}
\end{prop}

\proof For zero energy orbits we have
\[\frac{\d}{\d t}|x(t)|\leq \left|\dot x(t)\right|
\leq \sqrt{|2V(x(t))|}\leq C_1|x(t)|^{-\mu/2}.\]
This implies (\ref{eq:examlog00}) for large time.

Again, for zero energy orbits
\[\left|\dot x(t)\right|
=\sqrt{-2V(x(t))}\geq c_2x(t)^{-\mu/2},\ \ \  c_2>0,\]
 which together with (\ref{eq:examlog00}) yields 
(\ref{eq:velocity bound}) for  large time. For 
 positive energy orbits
we clearly have $|\dot x(t)|\to\sqrt{2\lambda}$, which also implies
(\ref{eq:velocity bound}) for large time.

Finally, (\ref{eq:alpha}) and
\[\left|\dot x(t)\right|
=\sqrt{-2V(x(t))}=O(|x(t)|^{-\mu/2})\]
yield 
(\ref{eq:examlog000}).
  \qed

 For a given outgoing scattering orbit $x(t)$ we
define the {
\it asymptotic normalized
  velocity} to be 
\begin{equation}
  \label{eq:asymptotic norm1}
 \omega^+=\lim_{t\to +\infty}\omega(t);\;\omega(t)= x(t)/|x(t)|,
\end{equation} provided that this limit exists.
We also define 
  \begin{equation*}
 \tilde\omega^+:=\lim_{t\to +\infty}\tilde\omega(t);\;
\tilde\omega(t)=\dot x(t)/|\dot x(t)|,  
  \end{equation*} 
provided that this limit exists.

\begin{prop}
  \label{thm:clas_vel} 
Suppose that
\begin{align}
\nabla^n V(x)&=O(|x|^{-n-\mu}),\ \  n=1,2,\nonumber
\\
 \ V(x)&\leq- c_0|x|^{-\mu},\ \
c_0>0,\ \ |x|\geq R,\nonumber\\
2V(x)+x\cdot \nabla V(x)&\leq -c|x|^{-\mu},\ \ \ c>0,\ \ \   |x|\geq
R,\nonumber\\ 
\nabla V(x)-\hat x \hat x \cdot\nabla
V(x)&=O(|x|^{-1-\mu-\epsilon_2}), \ \ \ \epsilon_2>0.
\label{connd}
\end{align} 
Then for any outgoing scattering orbit $x(t)$ there exists
$\omega^+$ and $\tilde\omega^+$ and they are equal.  Moreover,
\begin{equation}
  \label{eq:asymptotic norm2}
 \omega(t)=\omega^++O(t^{-{\alpha\epsilon_2}})=\tilde\omega(t)+O(t^{-{\alpha\epsilon_2}}).
\end{equation}
\end{prop}

\begin{proof}
Let $L_{ij}=x_i\dot x_j-x_j\dot x_i$ be the $ij$'th component of the
angular momentum. 
Note that
\[L^2:=\sum_{i<j}L_{ij}^2
=x^2\dot x^2-(x\cdot\dot x)^2
=x^2
\dot x^2(1-(\omega\cdot\tilde  \omega)^2).\]

By (\ref{connd}),
\begin{equation*}
  \left|\frac {\d}{\d t} L_{ij}\right|
=|x|\left|\nabla V(x)-\omega\, \omega\cdot\nabla V(x)\right|
=O(|x|^{-\mu-\epsilon_2})
=O\big (t^{-\alpha(\mu +\epsilon_2)}\big ),
\end{equation*}
and therefore, 
\begin{equation}
\label{eq:angular mom1}
  L_{ij}=O\big (t^{1-\alpha(\mu +\epsilon_2)}\big ).
\end{equation}

We compute
\[\frac{\d}{\d t}\omega(t)=\frac{\dot
  x(t)-\omega(t)\,\omega(t)\cdot\dot x(t)}{|x(t)|}.\]
Hence,
\begin{align*}
\left|\frac{\d}{\d t}\omega(t)\right|&=
\frac{\sqrt{\dot
  x^2(t)-
(\omega(t)\dot x(t))^2}}{|x(t)|}\\
&=\frac{|L(t)|}{|x(t)|^2}=O(t^{-1-\alpha\epsilon_2})\in L^1(\d t).
\end{align*}

Hence $\omega^+$ is well-defined, and the first estimate in
\eqref{eq:asymptotic norm2} holds.

Now
\begin{align*}
\frac{\d\tilde\omega(t)}{\d t}&=
-\frac{\nabla V(x(t))-\tilde\omega(t)\cdot\nabla V(x(t))\tilde\omega(t)}{|\dot
  x(t)|}\nonumber
\\
&=
-\frac{\omega(t)\cdot\nabla V(x(t))\big (\omega(t)
-\omega(t)\cdot\tilde\omega(t)\tilde\omega(t)\big )}{|\dot x(t)|}
+O(t^{-1-\alpha\epsilon_2}).\nonumber
\end{align*}
The norm of the first term equals
\[\frac{|\omega(t)\cdot\nabla V(x(t))||L(t)|}{|\dot
  x(t)|^2|x(t)|}=O(t^{-1-\alpha\epsilon_2}).\] 
Hence 
\[\frac{\d\tilde\omega(t)}{\d t}=
O(t^{-1-\alpha\epsilon_2})\in  L^1(\d t).\]
Hence $\tilde\omega^+$ is well-defined, and the second estimate in
\eqref{eq:asymptotic norm2} holds.

We have,
\[1-(\omega(t)\cdot \tilde\omega(t))^2=\frac{L(t)^2}
{x^2(t)\dot x^2(t)}=O(t^{-2\alpha\epsilon_2}).\]
Hence, $|\omega^+\cdot\tilde\omega^+|=1$.

By Proposition \ref{prop1} (or  \cite[(4.38)]{FS}), we have 
$\omega(t)\cdot\tilde\omega(t)\geq0$, for large $t$. Hence,
$\omega^+\cdot\tilde\omega^+\geq0$. Therefore,
$\omega^+=\tilde\omega^+$.
\end{proof}

\begin {example} \label{example:spiral}(Extension of an example in a preliminary version
  of the book \cite{DG}) Consider the potential $V=r^{-\mu}\chi(\theta-c\ln
  r)$ specified in two dimensions  using polar coordinates. Here $\chi \in C^{\infty}(S^1)$ is negative,
  $\chi'(0)<0$ 
  and $c>0$ (and $\mu\in (0,2)$). A computation shows that there is a
  classical scattering orbit with
  $\theta=c\ln r$ if 
  \begin{equation}
    \label{eq:examlog}
\chi(0)=\mu^{-1} \Big (c \big (\frac {\alpha}{1-\alpha\mu}-1\big )+c^{-1}\frac {1-\alpha}{1-\alpha\mu}\Big )\chi'(0).
  \end{equation}
So in this case the asymptotic normalized velocity $\omega^+$ does not
exist. This shows the importance of the smallness condition
of Condition \ref{assump:conditions1} 
(\ref{it:assumption4}), viz. $\epsilon_2>0$. We also see from \eqref{eq:examlog} that a
  weaker notion of smallness would not suffice neither: If $\chi$ is
  taken to be almost constant, say $\chi\approx -1$, which may be
  viewed as an example of another
  type of perturbed radial potential, then there is still a solution
  to  \eqref{eq:examlog}, in fact one with $c\approx0$.
\end{example}

\section{Mixed problem for radial potentials}
\label{Mixed problem in the case V_2= V_3= 0}

In this section we 
assume that
the potential is radial and for $r\geq 1$,
\begin{eqnarray}
&&|\partial_r^n V(r)|\leq C_n r^{-n-\mu},\ \ n=0,1,\dots;\nonumber\\
&&V(r)\leq -c r^{-\mu},\ \ c>0;\ \ rV'(r)+2V(r)< 0.
\label{piy}
\end{eqnarray}
Clearly, Condition \ref{assump:conditions1}
with $V_2(r)=0$ implies (\ref{piy}).

For radial potentials
 all orbits are confined to a plane. Let us first
investigate the two-dimensional problem.
We will use  polar coordinates
$(y_1,y_2)=(r\cos\theta,r\sin\theta)$.

 The angular momentum $L$ is a preserved quantity, at
  our disposal. We need to
  solve the system
\begin{equation}\label{eq:equations of motion}
\begin{cases}
\dot \theta=Lr^{-2}
\\
\dot r=\sqrt{2\lambda-2V(r)-L^2r^{-2}}
\end{cases}\;.
\end{equation}

We impose the conditions 
\begin{equation}
r(1)=r_1,\ \ \frac{\d }{\d t}r(1)>0,\ \ \lim_{t\to+ \infty}\theta(t)=0.
\label{conn}\end{equation}

Our assumption implies that for any $\lambda\geq0$ and  $L\in \R$,
there exists at most one $r_\crit=r_\crit(\lambda,L)\geq 1$ that solves
\[2\lambda-2V(r_\crit)-L^2r_\crit^{-2}=0.\]
Note that the function 
\begin{equation}
  \label{eq:incre}
  (1, \infty)\ni r\to rg(r) =r\sqrt{2\lambda-2V(r)}\text{ is increasing}.
\end{equation}

Clearly,
 for any $\lambda\geq0$, $L\in \R$ and $r_1>r_\crit\geq 1$ the problem
 \eqref{eq:equations of motion} subject to  (\ref{conn})
has a unique solution. This solution is a scattering orbit and it has  turning point at $r_\crit$.
 Writing  $\theta_1=\theta(1)$ we obtain
\begin{align}
  \label{eq:implicit cond}
 -\theta_1
&=L\int_{r_1}^{\infty}
  r^{-2}(2\lambda-2V(r)-L^2r^{-2})^{-\tfrac{1}{2}}\d r.
\end{align}
The angle between the asymptotic direction and the turning point equals
\begin{align}\label{eq:implicit cond22}
 -\theta_\crit
&=L\int_{r_\crit}^{\infty}
  r^{-2}(2\lambda-2V(r)-L^2r^{-2})^{-\tfrac{1}{2}}\d r.
\end{align}
Clearly,
 $\lim_{t\to-\infty}\theta(t)=2\theta_\crit$.

Let $\theta_\al=\theta_\al(\lambda,r_1)$ denote the largest allowed angle such that for
$|\theta_1|<\theta_\al$ there exists a solution to Newton's equation
with  energy $\lambda$ obeying the conditions 
(\ref{conn}) as well as $\theta(1)=\theta_1$.

\begin{prop}
Introduce the constant
\begin{equation}
  \label{eq:geometriccond}
  C=\sup_{r'\geq r\geq 1}\frac {V(r')}{V(r)}.
\end{equation}
Then $\theta_\al\geq\pi/2-\arctan\sqrt{C-1}$. 
In particular, if $V(r)$ is increasing, so that
$C=1$, then 
$\theta_\al\geq\pi/2$. 
\end{prop}

\begin{proof}
Let us write $L=r_1g(r_1) \kappa$ with $\kappa\in
[-1,1]$. 
It follows from \eqref{eq:incre} 
that for any such $\kappa$
\begin{equation*}
  2\lambda-2V(r)-L^2r^{-2}>0\;\text{for} \;r>r_1.
\end{equation*}
After a change of variable we may then write \eqref{eq:implicit cond}
as
\begin{equation}
  \label{eq:implicit2}
-\theta_1= \kappa  \int_1^\infty s^{-1}\left
 (s^2\frac{\lambda-V(sr_1)}{\lambda-V(r_1)}-\kappa^2\right)^{-\tfrac{1}{2}}\d
 s.
\end{equation}
Note that
\begin{equation}
cs^{-\mu}\leq\frac{\lambda-V(sr_1)}{\lambda-V(r_1)}=
\frac{g(sr_1)^2}{g(r_1)^2}\leq C.
\label{reinserted}\end{equation}

Clearly the right hand side of (\ref{eq:implicit2})
 is an increasing function of
$\kappa$. Therefore,
we get the lower bound 
\begin{eqnarray*}
\int_1^\infty s^{-1}\left
  (s^2\frac{\lambda-V(sr_1)}{\lambda-V(r_1)}-1\right)^{-\tfrac{1}{2}}\d
  s &\geq &\int_1^\infty s^{-1}\left (s^2C-1\right)^{-\tfrac{1}{2}}\d s\\
&=&\pi/2-\arctan \sqrt{C-1}
\end{eqnarray*} for the largest allowed 
angle. 
\end{proof} 

\begin{example} \label{example:Coulomb}
For the purely
  Coulombic case $V(r)=- \gamma r^{-1}$ one can compute 
the
  orbit\[L^2\gamma^{-1}r(t)^{-1}=1-\frac{\cos(\theta_\crit-\theta(t))}
{\cos(\theta_\crit)},\]
where $\theta_\crit(\lambda,L)=\pi-\arctan\sqrt{2\lambda L^2\gamma^{-2}}$,
 (see \cite[p. 126]{Ne}, for example). 
Therefore, the allowed angle equals
\begin{equation}\label{eq:Coulomb}
\theta_\al(\lambda,r_1)=
\pi-\arctan\sqrt{2\lambda(2\lambda \gamma^{-2}r_1^2+2\gamma^{-1}r_1)}.
\end{equation}
In particular, for $\lambda>0$ the
allowed angle is at least $\pi/2$ and for $\lambda=0$ it is $\pi$.
\end{example}

\begin{example} \label{example:hom-poten} We look at scattering for the
  example $V(r)=-\gamma r^{-\mu}$ at energy $\lambda=0$.
The angle betweeen asymptotic direction and the turning point
is independent of the orbit and is equal to
 $\theta_\crit=
\tfrac{\pi}{2-\mu}$. 
The fact that this angle is independent of the
  orbit may be seen independently by invoking the scaling and
  rotational symmetry of Newton's equation; thus there is essentially
  only one scattering orbit at  $\lambda=0$ (see the illustration in
  Section \ref{Introduction} 
  for the case $\mu=1.8$). 
The implicit equation for this orbit is
  \begin{equation}
    \label{eq:pol-eqn}{2\over 1+\cos\big( (2-\mu)(\theta_\crit-\theta(t))
\big)}=r(t)^{2-\mu}.   
  \end{equation} 
\end{example}

\subsection{Dependence of the angular momentum  on data}
\label{Dependence of kappa  on data}

\begin{lemma}
We fix  $\kappa_0\in (0,1)$.
Then for any $L\in \R$, $\lambda\geq0$ and $r_1\geq1$,
 satisfying 
\begin{equation}
\frac{L^2}{r_1^2}\leq \kappa_0^2g(r_1)^2,\label{yuy}
\end{equation}
 we have a unique
outgoing scattering orbit with the conditions
\eqref{conn}, the energy $\lambda$ and angular momentum $L$.
 The initial angle is given by \eqref{eq:implicit cond}, and we have the following estimates:
\begin{equation}
\partial_{r_1}^n\partial_{L^2}^m
\frac{\theta_1}{L} = O\left(
r_1^{-1-n-2m}g(r_1)^{-1-2m}\right),\ \ \ n,m\geq0.\label{esto}
\end{equation}
\end{lemma}

\begin{proof}
Only \eqref{esto} needs elaboration. For $n\geq1$, we have
\begin{align*}
\partial_r^n(2\lambda-2V(r)-L^2/r^2)&=O(r^{-\mu-n})+O(L^2r^{-2-n})\\
&= O(r^{-\mu-n})+O(r^2g(r)^2r^{-2-n})\\
&= O(r^{-n}g(r)^2).
\end{align*}
 The quantity $\partial_r^n(2\lambda-2V(r)-L^2/r^2)^{-p}
$  is a linear combination of terms of the following form, where
$n_1+\cdots+n_k=n$, 
\begin{align*}
(2\lambda-2V(r)-L^2/r^2)^{-p-k}
&\times\partial_r^{n_1}(2\lambda-2V(r)-L^2/r^2)
\cdots
\partial_r^{n_k}(2\lambda-2V(r)-L^2/r^2)\\
&=O\left(g(r)^{-2p-2k}
g(r)^2r^{-n_1}\cdots g(r)^2r^{-n_k}\right)\\
&=O\left(r^{-n}g(r)^{-2p}\right).
\end{align*}
Hence, 
\begin{equation*}
\partial_r^n(2\lambda-2V(r)-L^2/r^2)^{-p}
=O\left(r^{-n}g(r)^{-2p}\right).
\end{equation*}

Using \eqref{eq:incre} and \eqref{reinserted} we obtain that 
\begin{align*}\partial_{L^2}^m\frac{\theta_1}{L}&=C_m 
\int_{r_1}^\infty\d r r^{-2-2m}(2\lambda-2V(r)-L^2/r^2)^{-1/2-m}\\
&=O(r_1^{-1-2m}g(r_1)^{-1-2m}).
\end{align*}

For $n\geq1$, $\partial_{r_1}^n\partial_{L^2}^m\frac{\theta_1}{L}$
 is a  linear combination  of terms of the form
\begin{equation*}
  r_1^{-k-2m-1}\partial_{r_1}^{n-k}(2\lambda-2V(r_1)-L^2/r_1^2)^{-1/2-m} 
=O\big (r_1^{-n-2m-1}g(r_1)^{-1-2m}\big ).
\end{equation*}
\end{proof}

\begin{lemma}
Let $\theta_0\in (0,\pi/2-\arctan\sqrt{C-1})$ where $C$ is given by
\eqref{eq:geometriccond}. Then for all $r_1\geq 1$, $|\theta_1|\leq
\theta_0$  and $\lambda\geq 0$ we can find an outgoing
scattering orbit
satisfying \eqref{conn} with $\theta(1)=\theta_1$.
We have the following estimates:
\begin{align}
\partial_{r_1}^n \partial_{\theta_1^2}^m L^2&=O\left(
 r_1^{2-n} g(r_1)^{2}\right),\
 \
 n,m\geq0;
\label{estib2}\\
\partial_{r_1}^n \partial_{\theta_1^2}^m\frac{ L}{\theta_1}&= O\left(
 r_1^{1-n}g(r_1)\right),\ \
 n,m\geq0.
\label{estia2}
\end{align}
\label{pri}
\end{lemma}

\begin{proof}
 We can solve the equation \eqref{eq:implicit cond} for $L$ such that
 \eqref{yuy} is fulfilled for some $\kappa_0\in (0,1)$. Treating $L^2$ as
 an independent variable, obviously 
\begin{equation}
\partial_{r_1}^n \partial_{L^2}^m L^2=O\big (r_1^{2-2m-n}g(r_1)^{2-2m}\big ).
\label{hgh}\end{equation}
We apply 
$\partial_{r_1}^n \partial_{L^2}^m $ to
\[\theta_1^2=\left(\frac{\theta_1}{L}\right)^2L^2,\]
use (\ref{esto}) and (\ref{hgh}), and obtain
\begin{equation}
\partial_{r_1}^n \partial_{L^2}^m \theta_1^2=
O\big (r_1^{-n-2m}g(r_1)^{-2m}\big ).\label{estu}\end{equation}

Next we note
\begin{align}
\partial_{L^2}\theta_1^2 ={}&
\int_{r_1}^\infty r^{-2}(2\lambda-2V(r)-L^2r^{-2})^{-\frac12}\d r\nonumber\\
&\times\int_{r_1}^\infty
r^{-2}(2\lambda-2V(r))(2\lambda-2V(r)-L^2r^{-2})^{-\frac32}\d r \nonumber
 \\
\geq {} &c_0r_1^{-2}g(r_1)^{-2},\ \ \ \hbox{for some }c_0>0.
\label{esty}
\end{align}

We claim that the quantity $\partial_{r_1}^n \partial_{\theta_1^2}^m
L^2$ is a linear combination of terms of the form
\begin{equation}
\partial_{r_1}^{n_1}\partial_{L^2}^{m_1}\theta_1^2\cdots
\partial_{r_1}^{n_p}\partial_{L^2}^{m_p}\theta_1^2
(\partial_{L^2}\theta_1^2)^{-m-p}\label{formula}
=O\big (r_1^{2-n}g(r_1)^2\big ),
\end{equation} 
where $n=n_1+\cdots+n_p$ and 
 $m+p=m_1+\cdots+m_p+1$,
 which obviously proves (\ref{estib2}). 
To see  that indeed the terms are of the form given to  the left of (\ref{formula}) we use induction with respect to $n+m$.
The first step (justified by the implicit function theorem and the
chain rule)   is
\begin{align*}
\partial_{\theta_1^2}L^2&=(\partial_{L^2}\theta_1^2)^{-1},\\
\partial_{r_1}L^2&=-\partial_{r_1}\theta_1^2(\partial_{L^2}\theta_1^2)^{-1}.
\end{align*} 
The inductive step uses the following identities:
\begin{align*}
\partial_{\theta_1^2}(\partial_{r_1}^{n_1}\partial_{L^2}^{m_1}
\theta_1^2)
&=\partial_{r_1}^{n_1}\partial_{L^2}^{m_1+1}\theta_1^2
(\partial_{L^2}\theta_1^2)^{-1} 
,\\
\partial_{r_1}(\partial_{r_1}^{n_1}\partial_{L^2}^{m_1}\theta_1^2)
&=\partial_{r_1}^{n_1+1}\partial_{L^2}^{m_1}\theta_1^2
-\partial_{r_1}^{n_1}\partial_{L^2}^{m_1+1}\theta_1^2 \partial_{r_1}\theta_1^2
(\partial_{L^2}\theta_1^2)^{-1}.
\end{align*}
Finally we use (\ref{estu}) and (\ref{esty}) yielding the bound
(\ref{formula}).

The quantity $\partial_{r_1}^n \partial_{\theta_1^2}^m
\frac{\theta_1}{L}$ is a linear combination of terms of the form
\begin{equation*}
\partial_{r_1}^{n_1}\partial_{\theta_1^2}^{m_1}L^2\cdots
\partial_{r_1}^{n_p}\partial_{\theta_1^2}^{m_p}L^2
\partial_{r_1}^k\partial_{L^2}^p\frac{\theta_1}{L}
=O\big ( r_1^{-1-n}g(r_1)^{-1}\big )
,
\end{equation*}
where $n=n_1+\cdots+n_p+k$ and 
 $m=m_1+\cdots+m_p$; for the bound we
use  (\ref{esto}) and (\ref{estib2}). Thus
\begin{equation}
\partial_{r_1}^n \partial_{\theta_1^2}^m \frac{\theta_1}{L}
=O\big ( r_1^{-1-n}g(r_1)^{-1}\big ).
\label{este}
\end{equation}
We note the inequality
\begin{equation}
-\frac{\theta_1}{L}\geq r_1^{-1}g(r_1)^{-1}.
\label{esta}\end{equation}

Finally, the quantity $\partial_{r_1}^n \partial_{\theta_1^2}^m\frac{
  L}{\theta_1}$ is  a linear
combination of terms of the form
\begin{equation*}
\partial_{r_1}^{n_1}\partial_{\theta_1^2}^{m_1}\frac{\theta_1}{L}\cdots
\partial_{r_1}^{n_k}\partial_{\theta_1^2}^{m_k}\frac{\theta_1}{L}
\left(\frac{L}{\theta_1}\right)^{k+1}
=O( r_1^{1-n}g(r_1))
,
\end{equation*} 
where 
$n_1+\cdots+n_k=n$ and  $m_1+\cdots+m_k=m$; for the bound  
 we use (\ref{este}) and (\ref{esta}). This proves
(\ref{estia2}).
\end{proof}

Finally, we consider  orbits  in an arbitrary dimension.
We introduce for $R\geq 1$ and $\sigma>0$
\begin{align*}
  \Gamma^+_{R,\sigma}(\omega)&=\{y\in {\mathbb R}^d\;|\;y\cdot
  \omega\geq (1-\sigma)|y|,\;|y|\geq R\};\; \omega\in S^{d-1},\\ 
  \Gamma^+_{R,\sigma}&=\{(y,\omega)\in {\mathbb R}^d \times
  S^{d-1}|\;y \in \Gamma^+_{R,\sigma}(\omega)\}.
 \end{align*}
The mixed problem  consists in finding a solution $y(t)$ to
Newton's equation subject to mixed boundary conditions and an energy
 constraint given in terms of data $x\in {\mathbb R}^d,\;
\omega\in S^{d-1}$ and $\lambda\geq 0$:
\begin{equation}\label{eq:mixed conditions}
\begin{cases}
\ddot y(t) =-\nabla V(y(t)),\\
\lambda=\tfrac {1}{2 }\dot y(t)^2 +V(y(t)),\\
y(1)=x,\\
\omega=\lim_{t\to +\infty}\omega(t);\;\omega(t)= y(t)/|y(t)|.
\end{cases}\;
\end{equation}

\begin{prop} \label{lemma:mixed_1}
For all  
small enough $\sigma>0$ the problem \eqref{eq:mixed conditions} has a
 solution $y(t),\;t\geq 1$, for all data 
$(x,\omega)\in \Gamma^+_{1,\sigma}$ and $\lambda \geq 0$. Moreover this
solution $y(t)\in \Gamma^+_{1,\sigma}(\omega)$ for all $t\geq 1$, and 
given the latter invariance property it is  unique and
\begin{eqnarray}
\partial_{x}^\alpha \partial_{\omega}^\beta
 L^2&=& O\left (|x|^{2-|\alpha|}g(|x|)^2\right),
\label{esti3}
\end{eqnarray}
\end{prop}

\begin{proof}
 Note that $(r_1,\sin^2\theta_1)=\big (|x|,1-(\omega \cdot \hat x)^2\big )$.
Therefore, $\theta_1^2$
and $r_1=|x|$ are smooth function of $x$ and $\omega$ with
\begin{equation}\label{eq:mixed_1as2}
  \partial_\omega^\delta \partial_x^\gamma\theta_1^2=O(| x
|^{-|\gamma|}),\;\partial_x^\gamma r_1 =O(| x
|^{1-|\gamma|}).\end{equation} 
In conjunction with \eqref{estib2} and the Faa di Bruno formula we
obtain \eqref{esti3}. \end{proof}
\begin{remarks}
 \label{remark:nonsmooth} \begin{enumerate}[\quad\normalfont 1)]
\item \label{it:konkap1} The function $L^2$ is   continuous in all
  variables  at $\lambda=0$,  however as
  may readily be checked it is  not smooth in $\lambda$
  at this point. This function is smooth for $\lambda>0$.
 \item \label{it:konkap2} The derivatives in $\omega$
  and $x$ of the  function $L^2$ are also
  continuous in $\lambda$ at $\lambda=0$. This follows by an abstract
  argument (the
  proof is very simple, see for example \cite[proof of Lemma
  7.7.2]{Ho1}): Suppose $\mathcal U$ is an open subset of $\R^n$, and
  that $f:\mathcal U\times [0,1]\to \R$ is smooth in $z\in\mathcal U$
  (for any fixed $\lambda \in [0,1]$) with $|\partial_z^\beta f|\leq
  C_\beta$ uniformly on $\mathcal U\times [0,1]$, and suppose $f$ 
  is continuous in $(z,\lambda)\in\mathcal U\times [0,1]$. Then 
  all $z$--derivatives are  continuous in $(z,\lambda)\in\mathcal U\times
  [0,1]$.
\end{enumerate} 
\end{remarks}

\subsection{Dependence of flow on data}
\label{Dependence of flow on data}
Let us examine the dependence of 
the flow on the boundary conditions. We start with the dependence
of the two-dimensional flow
$(\theta,r)=(\theta(t),r(t))$ on $(r_1,\theta_1)$.

\begin{lemma}  \label{thm:properties}The orbits described in 
Lemma  \ref{pri} obey 
\begin{align}
\partial_{r_1}^n\partial_{\theta_1^2}^mr&=O\left(r_1^{1-n}
g(r_1)  g(r)^{-1}  \right);\ 
n+m\geq1, 
\label{est1}\\
\partial_{r_1}^n\partial_{\theta_1^2}^m \theta^2&=
O\left(r_1^{2-n}r^{-2}
g(r_1)^{2}  g(r)^{-2}  \right);\ n+m\geq0,\label{est2}\\
\partial_{r_1}^n\partial_{\theta_1^2}^m \frac{\theta}{\theta_1}&=
O\left(r_1^{1-n}r^{-1}
g(r_1)  g(r)^{-1}  \right);\ n+m\geq0.\label{est3}
\end{align}
\end{lemma}

\begin{proof}
To prove (\ref{est1}) we note that
 the second equation of \eqref{eq:equations of motion} is
solved by 
\begin{equation}
  \label{eq:solution for r}
  \int_{r_1}^r (2\lambda-2V(\rho)-L^2\rho^{-2})^{-1/2}\d 
\rho=t-1.
\end{equation}
We use  induction wrt. $n+m$.
We apply to (\ref{eq:solution for r}) the derivative
$\partial_{r_1}^n\partial_{\theta_1^2}^m$. We obtain that zero is 
a linear combination of terms of the following form:
\begin{equation}
\partial_{r_1}^{n_1}\partial_{\theta_1^2}^{m_1}L^2\cdots
\partial_{r_1}^{n_k}\partial_{\theta_1^2}^{m_k}L^2
\times r_1^{-2k-u}\partial_{r_1}^v (2\lambda-2V(r_1)-L^2/r_1^2)^{-\frac12-k},
\label{term1}\end{equation}
with $n_1+\cdots +n_k+u+v+1=n$, $m_1+\cdots+m_k=m$;
\begin{align}
\partial_{r_1}^{p_1}\partial_{\theta_1^2}^{q_1}r\cdots
\partial_{r_1}^{p_l}\partial_{\theta_1^2}^{q_l}r &\times
\partial_{r_1}^{n_1}\partial_{\theta_1^2}^{m_1}L^2\cdots
\partial_{r_1}^{n_k}\partial_{\theta_1^2}^{m_k}L^2 \nonumber
\\
&\times
r^{-2k-u}\partial_{r}^v (2\lambda-2V(r)-L^2/r^2)^{-\frac12-k},
\label{term2}\end{align}
with  $n_1+\cdots +n_k+p_1+\cdots+p_l=n$,
$u+v+1=l$, $m_1+\cdots+m_k+q_1+\cdots+q_l=m$;
and
\begin{equation}
\partial_{r_1}^{n_1}\partial_{\theta_1^2}^{m_1}L^2\cdots
\partial_{r_1}^{n_k}\partial_{\theta_1^2}^{m_k}L^2 \times
\int_{r_1}^r\rho^{-2k}
(2\lambda-2V(\rho)-L^2/\rho^2)^{-\frac12-k}\d\rho,
\label{term3} 
\end{equation}
with $n_1+\cdots +n_k=n$, $m_1+\cdots+m_k=m$.

Using \eqref{estib2} the terms (\ref{term1}) are estimated  by
\begin{equation}
O\left(r_1^{-n+1}g(r_1)^{-1}\right).
\end{equation}

The terms (\ref{term2}) are divided into the single term
\begin{equation}
\partial_{r_1}^n\partial_{\theta_1^2}^m r\ (2\lambda-2V(r)-L^2/r^2)^{-1/2}
\end{equation}
and the remaining ones, which by \eqref{eq:incre} and \eqref{estib2} can be estimated by
\begin{equation}
C|\partial_{r_1}^{p_1}\partial_{\theta_1^2}^{q_1}r|
\cdots
|\partial_{r_1}^{p_l}\partial_{\theta_1^2}^{q_l}r|
r_1^{p_1+\cdots+p_l-n}r^{1-l}g(r)^{-1}.\label{term4}
\end{equation}
By the induction assumption, and using $l\geq1$ and \eqref{reinserted},  (\ref{term4})
is bounded by
\begin{equation}
Cr_1^{1-n}g(r_1) g(r)^{-2}.
\end{equation}

Using $k\geq1$ and \eqref{reinserted}, the terms (\ref{term3}) are
estimated by 
\begin{align}
C_1
g(r_1)^{2k}r_1^{2k-n}\int_{r_1}^r\rho^{-2k}
g(\rho)^{-1-2k}\d\rho
&\leq C_2g(r_1)^{2}g(r)^{-2} r_1^{2-n}
\int_{r_1}^r\rho^{-2}g(\rho)^{-1}\d\rho
\nonumber
\\
&\leq
C_3g(r_1)g(r)^{-2} r_1^{1-n}.
\end{align}
Thus we obtain the  estimate
\begin{equation}
|\partial_{r_1}^n\partial_{\theta_1^2}^m
r\,|g(r)^{-1} 
=O\left(g(r_1)g(r)^{-2}r_1^{1-n}\right),
\end{equation}
from which (\ref{est1}) follows.

Next we would like to prove (\ref{est2}). We start from the identity
\begin{equation}
\frac{\theta}{L}
=-\int_{r}^\infty\rho^{-2}(2\lambda-V(\rho)-L^2/\rho^2)^{-\frac12}\d\rho.
\label{equi}\end{equation}
This shows $\frac{\theta}{L}=O\big (r^{-1}g(r)^{-1}\big )$. Next
we obtain that
$\partial_{r_1}^n\partial_{\theta_1^2}^m\frac{\theta}{L}$ 
 is a linear combination of terms of the following form:
\begin{align}
\partial_{r_1}^{p_1}\partial_{\theta_1^2}^{q_1}r\cdots
\partial_{r_1}^{p_l}\partial_{\theta_1^2}^{q_l}r &\times
\partial_{r_1}^{n_1}\partial_{\theta_1^2}^{m_1}L^2\cdots
\partial_{r_1}^{n_k}\partial_{\theta_1^2}^{m_k}L^2 
\nonumber\\
&\times
r^{-2-2k-u}\partial_{r}^v (2\lambda-2V(r)-L^2/r^2)^{-\frac12-k},
\label{term1a}
\end{align}
with  $n_1+\cdots +n_k+p_1+\cdots+p_l=n$,
$u+v+1=l$, $m_1+\cdots+m_k+q_1+\cdots+q_l=m$;
and
\begin{equation}
\partial_{r_1}^{n_1}\partial_{\theta_1^2}^{m_1}L^2\cdots
\partial_{r_1}^{n_k}\partial_{\theta_1^2}^{m_k}L^2 \times
\int_{r}^\infty\rho^{-2-2k}
(2\lambda-2V(\rho)-L^2/\rho^2)^{-\frac12-k}\d\rho,
\label{term2a} 
\end{equation}
with $n_1+\cdots +n_k=n$, $m_1+\cdots+m_k=m$.

The term (\ref{term1a}) is estimated by
\begin{equation}
C_1 r_1^{-n+l+2k}r^{-1-l-2k}
g(r_1)^{l+2k}g(r)^{-1-l-2k}
\leq 
C_2r_1^{-n}r^{-1}
g(r)^{-1}.
\end{equation}

The term (\ref{term2a}) is estimated by
\begin{equation}
C_1 r_1^{-n+2k}r^{-1-2k}g(r_1)^{2k}g(r)^{-1-2k}
\leq 
C_2r_1^{-n}r^{-1}
g(r)^{-1}
.\end{equation}
Thus
\begin{equation}
\partial_{r_1}^n\partial_{\theta_1^2}^m\frac{\theta}{L}
=O\left(r_1^{-n}r^{-1}
g(r)^{-1}\right) .
\label{est4}
\end{equation}

Now by 
\[
\theta^2=
\left(\frac{\theta}{L}\right)^2 L^2,
\]
(\ref{estib2})  and (\ref{est4})  we obtain (\ref{est2}).

By
\[\frac{\theta}{\theta_1}=\frac{\theta}{L}\frac{L}{\theta_1},\]
(\ref{estia2}) and (\ref{est4}), we obtain  (\ref{est3}).
\end{proof}

We go back to the case of an arbitrary dimension.

\begin{prop}
The orbits considered in Proposition   \ref{lemma:mixed_1} satisfy
\begin{equation}
\partial_\omega^\delta \partial_x^{\gamma} y =
\begin{cases}
  O(|y|)&\text { for }\gamma=0;\\
 O\left(|x|^{1-|\gamma|}
g(|x|)g(|y|)^{-1}
\right)&\text { for }|\gamma|\geq 1;
\end{cases} \label{eq:y_x_der}
 \end{equation} in particular,
\begin{equation}
  \label{eq:y_x_derhigher}
 \partial_\omega^\delta \partial_x^{\gamma} y =O \big (|x|
 ^{-|\gamma|}|y|\big ). 
\end{equation} 
\end{prop}

\begin{proof}
We use
 the formula
\begin{equation}\label{eq:yformu1}
  y=r\cos \theta \;\omega +r\frac{\sin \theta}{\sin \theta_1}(\hat
  x-\hat x\cdot\omega\;\omega).
\end{equation}
Now, $\partial_\omega^\delta\partial_x^\gamma  r\cos\theta\omega$
is a linear combination of terms of the form
\begin{multline}\label{jjess1}
\partial_\omega^{\pi_1}\partial_x^{\rho_1}r_1\cdots
\partial_\omega^{\pi_n}\partial_x^{\rho_n}r_1
\times
\partial_\omega^{\sigma_1}\partial_x^{\tau_1}\theta_1^2\cdots
\partial_\omega^{\sigma_m}\partial_x^{\tau_m}\theta_1^2
\\
\times
\partial_{r_1}^{n_1}\partial_{\theta_1^2}^{m_1}
r\partial_{r_1}^{n_2}\partial_{\theta_1^2}^{m_2} \cos\theta 
\times\partial_\omega^{\delta_0}\omega
.
\end{multline}
Likewise,
$
\partial_\omega^\delta\partial_x^\gamma 
\frac{r\sin\theta}{r_1\sin\theta_1}(x-x\cdot\omega\ \omega)$
is a linear combination of terms of the form
\begin{multline}\label{jjess2}
\partial_\omega^{\pi_1}\partial_x^{\rho_1}r_1\cdots
\partial_\omega^{\pi_n}\partial_x^{\rho_n}r_1
\times
\partial_\omega^{\sigma_1}\partial_x^{\tau_1}\theta_1^2\cdots
\partial_\omega^{\sigma_m}\partial_x^{\tau_m}\theta_1^2
\\
\times
\partial_{r_1}^{n_1}\partial_{\theta_1^2}^{m_1} 
r \partial_{r_1}^{n_2}\partial_{\theta_1^2}^{m_2} \frac{\sin
  \theta}{\sin \theta_1} 
\times
\partial_\omega^{\delta_0}\partial_x^{\gamma_0}(\hat x-
\hat x\cdot\omega\ \omega)
.\end{multline}

Note that
\begin{align}\label{jess1}
\partial_\omega^{\delta_0} \omega &=
O(1),\\
\label{jess2}\partial_\omega^{\delta_0}\partial_x^{\gamma_0} 
(\hat x-\hat x\cdot\omega\;\omega)&=O(|x|^{-|\gamma_0|}).
\end{align}

Moreover by Lemma  \ref{thm:properties},
\begin{align}
\partial_{r_1}^{n_1}\partial_{\theta_1^2}^{m_1}r
&=
\begin{cases}
O(r),&n_1+m_1=0,\\
r_1^{1-n_1}g(r_1)  g(r)^{-1},&
n_1+m_1\geq1;
\end{cases}\label{ess1}
\\
\partial_{r_1}^{n_2}\partial_{\theta_1^2}^{m_2} \cos\theta 
&=
\begin{cases}
O(1),&n_2+m_2=0\\
r_1^{2-n_2}r^{-2}g(r_1)^2  g(r)^{-2},&
n_2+m_2\geq1;
\end{cases}\label{ess2}\\
\partial_{r_1}^{n_2}\partial_{\theta_1^2}^{m_2}\frac{ \sin\theta
}{\sin\theta_1} 
&=
r_1^{1-n_2}r^{-1}g(r_1)  g(r)^{-1}.
\label{ess4}
\end{align}
(For \eqref{ess4} we use the decomposition $\frac{ \sin\theta
}{\sin\theta_1} =\frac{\theta}{\theta_1}\frac{ \sin (\theta)/\theta
}{\sin(\theta_1)/\theta_1}$.)

Now, applying \eqref{eq:mixed_1as2} and \eqref{jess1}--\eqref{ess4} to
\eqref{jjess1} and \eqref{jjess2} yields \eqref{eq:y_x_der}.
\end{proof}

One may also estimate derivatives of ${ \dot y}$:
\begin{prop}
\begin{equation}
  \label{eq:y_x_derhigher2}
 \partial_\omega^\delta \partial_x^{\gamma}(\dot  y -\sqrt{2\lambda}\omega)=
O\big( |x|^{-|\gamma|}|y|^{-\mu} g(|y|)^{-1} \big ). 
\end{equation}

In particular
 \begin{equation}
  \label{eq:y_x_derhigher222f}
 \partial_\omega^\delta \partial_x^{\gamma}\dot  y=
O\big( |x|^{-|\gamma|} g(|y|)\big ). 
\end{equation}
\end{prop}

\begin{proof}
 First we
represent 
\begin{equation}
  \label{eq:F_rep}
  \dot y (t)-\sqrt {2\lambda}\omega=\int _t^\infty \nabla V(y)\d t',
\end{equation}
Now
$\partial_\omega^\delta\partial_x^\gamma
(  \dot y (t)-\sqrt {2\lambda}\omega)$ is a linear combination of terms of the
form
\begin{align}
&\int_t^\infty\partial
_\omega^{\delta_1}\partial_x^{\gamma_1} y(t')\cdots
\partial
_\omega^{\delta_n}\partial_x^{\gamma_n} y(t')
\nabla^{n+1}V(y(t'))\d t'\nonumber\\
&\qquad=O\left(|x|^{-|\gamma|}
\int_{|y|}^\infty
\rho^{-1-\mu}g(\rho)^{-1}\d\rho,\right)
\label{eq:dot y_x_der}, 
\end{align}
which are $O\big( |x|^{-|\gamma|}|y|^{-\mu}g(|y|)^{-1}\big
)$.  
\end{proof}

We also notice the uniform bounds
\begin{equation}
  \label{eq:g/r-bound}
 |y|^{-1}g(|y|)\leq \tfrac {C}{t-1},\; |y|^{-1}\leq Ct^{-\alpha},
\end{equation}
Since $|y|^{-1}\leq |x|^{-1}$ the second estimate of
\eqref{eq:g/r-bound} may be generalized as 
\begin{equation}
  \label{eq:generalized r-bound}
|y|^{-1}\leq C|x|^{-\delta}t^{-\alpha(1-\delta)};\;\delta \in [0,1].  
\end{equation}

\section{Time--dependent linear force problem}
\label{Time--dependent linear force problem }

We consider the following one-dimensional matrix-valued ODE
\begin{equation}
  \label{eq:z_1dim}
 \ddot z(t) -q(t)z(t)=\tilde z(t),\; t\geq 1,
\end{equation} where $q(t)\in M_d(\C)$ is self-adjoint for all $t\geq
1$, and as a function of $t$, $q$ is continuous and bounded. Moreover we assume
the following bound for some $\epsilon>0$
\begin{equation}
  \label{eq:q(t)}
(t-1)^2q(t)\geq -4^{-1}(1-\epsilon^2)\;\text {for}\;t\geq 1.
\end{equation}

The goal of this section is to 
study the initial value problem given by \eqref{eq:z_1dim} and the
initial value condition $z(1)=0$. As the reader will see  the  relevant tools come from  functional
analysis. 

Throughout the section we fix any 
\begin{equation}
  \label{eq:svalue}
  r\in (-\epsilon/2,\epsilon/2).
\end{equation}

We introduce for any such $r$ the following form on the domain
$\mathcal D(Q_r)=W^{1,2}_0(1,\infty)\allowbreak \subseteq L^2(1,\infty)$
($W^{1,2}_0\subseteq  W^{1,2}$ refer to  standard Sobolev spaces, see
for example \cite{D}, although we are here dealing with $\C^d$--valued
functions). Let 
$p_t=-\i {\d \over \d t}$.
\begin{equation}
  \label{eq:form_s}
  Q_r(\psi,\phi)=\langle p_t \psi,  p_t \phi\rangle +\langle \psi,
  \{q(t)-r^2t^{-2}+\i r(p_tt^{-1}+t^{-1}p_t)\}\phi\rangle.
\end{equation}

Formally this is the form of the operator $H_r=t^rHt^{-r}$, where $H$
is the Schr\"odinger operator $H=
p_t^2+q(t)$ with Dirichlet boundary condition at $t=1$. To justify this we invoke \cite[Theorem VIII.16]{RS}  and
the Hardy inequality 
\cite[Lemma 5.3.1]{D}. Due to
this  inequality and 
\eqref{eq:q(t)} there exists $\delta=\delta(\epsilon,r)>0$ such that 
\begin{equation}
  \label{eq:eqss}
 \Re  Q_r(\phi)= \Re  Q_r(\phi,\phi)\geq \delta\langle \phi,\{p_t^2+(t-1)^{-2}\}\phi\rangle.
\end{equation} Whence, in the terminology of  \cite{RS} the form $Q_r$
is strictly $m$--accretive. There is an associated operator $H_r$ for
which the open left half-plane $\C_-=\{\zeta\in \C|\Re \zeta<0\}$ is a subset
of the resolvent set,
cf. \cite[Lemma after Theorem VIII.16]{RS}. 

\begin{lemma}
\label{lemma:reso1}  The $\mathcal B (L^2(1,\infty))$--valued functions
\begin{equation*}
 B_r(\zeta):=(t-1)^{-1}(H_r-\zeta)^{-1}(t-1)^{-1} 
\end{equation*} 
and
\begin{equation}
  \label{eq:p-t}
  p_t(H_r-\zeta)^{-1}(t-1)^{-1} 
\end{equation}
 are uniformly bounded on $\C_-$. 
\end{lemma}
\begin{proof}
  Applying \eqref{eq:eqss} to $\phi=(H_r-\zeta)^{-1}(t-\sigma)^{-1}f$
  where $\sigma<1$ and $f\in L^2(1,\infty)$ yields (by the Cauchy
  Schwarz inequality) 
  \begin{equation}\label{eq:resbnd1}
    \|(t-\sigma)^{-1}(H_r-\zeta)^{-1}(t-\sigma)^{-1}f\|\leq \delta^{-1}\|f\|.
  \end{equation} Letting $\sigma\to 1$ using the Lebesgue  convergence theorems we
  conclude
that $(H_r-\zeta)^{-1}(t-1)^{-1}f\in \mathcal D((t-1)^{-1})$ for all
$f\in \mathcal D((t-1)^{-1})$  and that \eqref{eq:resbnd1} with
$\sigma=1$ holds for such $f$'s.

As for bounding \eqref{eq:p-t} we combine \eqref{eq:eqss} and \eqref{eq:resbnd1}
to obtain uniform boundedness of
$p_t(H_r-\zeta)^{-1}(t-\sigma)^{-1}$. Again we let $\sigma\to 1$.
\end{proof}

\begin{lemma}
\label{lemma:reso2}  There exists the weak 
limit 
\begin{equation*}
 B_r(0)=\w-\lim _{\zeta\to 0,\,\Re \zeta<0} B_r(\zeta). 
\end{equation*}
\end{lemma}
\begin{proof} Let $f\in L^2(1,\infty)$. For any sequence $\zeta_n\to 0$
  with $\Re
  \zeta_n<0$, $B_r(\zeta_{n_k})f\rightharpoonup g$ for some $g$ and some
  subsequence  $\zeta_{n_k}$, c.f. Lemma \ref{lemma:reso1} and
  \cite[Theorem V.2.1]{Yo}. Writing  $g=B_r(0)f$ it remains to
  show that for any $f\in \mathcal D((t-1)^{-1})$, $g$ is
  independent of choice of sequences. So suppose that for such $f$, 
  $B_r(\zeta_{1,n})f\rightharpoonup g_1$ and
  $B_r(\zeta_{2,n})f\rightharpoonup g_2$. We need to show that 
  $\psi:=g_1-g_2=0$. We readily obtain  that
  for all $\phi\in C^\infty_c(1,\infty)$
  \begin{equation}\label{eq:resbnd111}
    \langle (t-1)H_{-r}\phi,\psi \rangle=0. 
  \end{equation} Using this we can show that indeed $\psi=0$ by the
  following approximation argument. Pick a real-valued $\chi \in C^\infty[1,\infty)$ such
  that $\chi(t)=1$ for $t<2$ and $\chi(t)=0$ for $t>3$, and  let
  $\chi_n(t)=\chi(t/n)$; $n\in \N$. We introduce 
  $\Psi(t)=(t-1)\psi(t)$ and $\psi_n=\chi_n\Psi$. By elliptic regularity we obtain
  from \eqref{eq:resbnd111} that $\Psi(t)$ is
  smooth up to (and including) $t=1$. Since $\psi\in L^2(1,\infty)$ we must have
  $\Psi(1)=0$, in particular $\psi_n\in \mathcal D
  (Q_r)$. Using  \eqref{eq:resbnd111} with this input we 
  compute
\begin{equation}\label{eq:resbnd1111}
  Q_r(\psi_n)= \|\chi_n'\Psi\|^2.
\end{equation} 

Since $(t-1)\chi_n'(t)=(t-1)\chi'(t/n)/n$ is uniformly bounded, the Lebesgue
dominated convergence theorem yields that the right hand side of
\eqref{eq:resbnd1111} vanishes as $n\to \infty$. Whence by
\eqref{eq:eqss} 
\begin{equation*}
 \|\psi\|=\lim_{n\to \infty}\|\chi_n\psi\|=0. 
\end{equation*}
\end{proof}

\begin{lemma}
\label{lemma:reso3}  For all $\zeta  \in \C_-$ 
\begin{equation*}
 (H_r-\zeta)^{-1}=t^{r}(H-\zeta)^{-1}t^{-r}.
\end{equation*}
\end{lemma}
\begin{proof} We shall only consider the case $r\geq 0$ (the case 
  $r\leq 0$ may be treated similarly). It suffices to show that 
  \begin{equation}\label{eq:resbnd11102}
    t^{-r}(H_r-\zeta)^{-1}=(H-\zeta)^{-1}t^{-r}.
  \end{equation} Clearly $t^{-r}(H_r-\zeta)^{-1}f\in \mathcal D
  (Q_0)$ for any $f\in L^2(1,\infty)$. For all $\phi\in
  C^\infty_c(1,\infty)$ we compute 
  \begin{equation}\label{eq:resbnd1110}
    Q_0(\phi,t^{-r}(H_r-\zeta)^{-1}f)=\langle \phi,
    t^{-r}H_r(H_r-\zeta)^{-1}f\rangle. 
  \end{equation} Since $C^\infty_c(1,\infty)$ is a core for $Q_0$ we
    deduce that \eqref{eq:resbnd1110} is valid for all $\phi\in \mathcal D
  (Q_0)$. Whence  
  \begin{equation*}
    Ht^{-r}(H_r-\zeta)^{-1}f=t^{-r}H_r(H_r-\zeta)^{-1}f,
  \end{equation*} from which we readily obtain \eqref{eq:resbnd11102}.
\end{proof}

Using Lemma \ref{lemma:reso3} we can show strong convergence 
of 
$T_r(\zeta):=t^{-1}(H_r-\zeta)^{-1}t^{-1}$. Notice that by  Lemma \ref{lemma:reso2}
  there exists $$T_r(0)=\w-\lim _{\zeta\to 0,\,\Re \zeta<0} T_r(\zeta),$$
  and that by Lemma \ref{lemma:reso3}, $T_r(0)=t^rT_0(0)t^{-r}$. 

\begin{lemma}
\label{lemma:reso4}  
\begin{equation}\label{eq:gconv0}
 T_r(0)=\s-\lim _{\zeta\to 0,\,\Re \zeta<0} T_r(\zeta). 
\end{equation}
\end{lemma}
\begin{proof}
  Pick $\delta\in (0,\epsilon/2-|r|)$. We
 claim that for
 all $f\in \mathcal D (t^\delta)$ 
 \begin{equation}
   \label{eq:gconv}
  T_r(\zeta)f=t^{-\delta}T_{r+\delta}(\zeta)t^\delta f\to t^{-\delta}T_{r+\delta}(0)t^\delta f.
 \end{equation} 

Since $t^{-\delta}t^{-1}(H_r+1)^{-1}t$ is
 compact the (norm-) convergence \eqref{eq:gconv} follows from the first resolvent equation and
 weak convergence. Since $t^{-\delta}T_{r+\delta}(0)t^\delta
 =T_{r}(0)$ this proves \eqref{eq:gconv0}.
\end{proof}

\begin{remark}
 \label{remark:reso5} Using the uniform boundedness of the family \eqref{eq:p-t}
  and Lemma \ref{lemma:reso2} one may show that there
 exists the weak limit 
 \begin{equation*}
  \w-\lim _{\zeta\to 0,\,\Re \zeta<0}
 R_r(\zeta); \;R_r(\zeta)=p_t(H_r-\zeta)^{-1}t^{-1}. 
 \end{equation*} Assuming in addition to the given
 conditions on $q$ that $tq(t)$ is bounded, 
 one may show using the
 proof of Lemma \ref{lemma:reso4}  that there
 exists the strong limit, $\s-\lim _{\zeta\to 0,\,\Re \zeta<0} R_r(\zeta)$.   
\end{remark}

We  introduce for  $s \in \R$  the weighted spaces
\begin{equation}
  \label{eq:space}
  Z_{-s}=L^2_{-s}(1,\infty)=t^s L^2(1,\infty);
\end{equation}  here  $L^2$ refers to the space of $\C^d$--valued square integrable functions.

\begin{lemma}
\label{lemma:reso333} Suppose $s<1+\tfrac{\epsilon}{2}$, where
 $\epsilon>0$ is given as in \eqref{eq:q(t)}. Suppose $z\in
 L^2_{-s}(1,\infty)$ satisfies the homogeneous analogue of
 \eqref{eq:z_1dim} in $\mathcal
 D'(1,\infty)$ (i.e. we assume that the right hand side vanishes and
 that the equation holds in distributional sense), and $z(1)=0$. Then $z=0$.
\end {lemma}
\begin{proof}
  We consider for any $\tilde z \in \mathcal D (1,\infty)$ 
  \begin{equation*}
    0=\lim_{\zeta\to 0,\,\Re \zeta<0}\langle (H-\zeta)^{-1} \tilde z, Hz\rangle. 
  \end{equation*} By Lemmas \ref{lemma:reso3} and \ref{lemma:reso4} we
  may compute the limit  as to obtain
\begin{equation*}
    0=\langle H^{-1}\tilde z, Hz\rangle
  \end{equation*} with $H^{-1}\tilde z:=t^{1-r}T_r(0)t^{1+r}\tilde z$
  provided  $|r|<\epsilon/2$. For a later application we need 
  $r\geq s-1$ which by assumption is feasible. 

The idea is to integrate by parts in the
  expression to the right. First we notice that $p_tH^{-1}\tilde z\in
  L^2_{r}$, c.f. Remark \ref{remark:reso5}. Next we claim that 
  \begin{equation}
    \label{eq:-delta+1}
    p_tz\in L^2_{1-s}.
  \end{equation} For that we introduce for $n\in \N$ the
  multiplication operator $F_n(t)=F(t/n<1)$, and consider the expression
  \begin{equation} \label{eq:-delta+2}
 \langle p_tz, F_n(t) t^{2-2s}p_tz\rangle+\langle z, q(t)F_n(t) t^{2-2s}z\rangle.    
  \end{equation} Up to a term that can be bounded uniformly in $n$
  (using the assumption that $z\in
 L^2_{-s}$) this expression is equal to $\Re \langle Hz, F_n(t)
 t^{2-2s}z\rangle$. Whence \eqref{eq:-delta+1} follows from
 \eqref{eq:-delta+2} and the monotone convergence theorem.

Now  integration by parts yields (this is a version of Green's identity)
\begin{equation*}
    0=\Big [-\overline {H^{-1}\tilde z}\cdot {\d \over \d t}z+\Big \{{\d
    \over \d t}\overline  {H^{-1}\tilde
    z}\Big \}\cdot z\Big ]^\infty_1+\langle HH^{-1}\tilde z, z\rangle.
  \end{equation*} Obviously the last term to the right is equal to
  $\langle \tilde z, z\rangle$. We claim that the first term
  vanishes. The lower boundary term  (at  $t=1$)  vanishes. The upper
  limit should be interpreted as a limit along a suitable sequence
  $t_m\to \infty$. Specifically, since the form is
  $[tf(t)]^\infty_1$  with $f$ integrable (here we use
  \eqref{eq:-delta+1}) indeed $t_mf(t_m)\to 0$
  along such  sequence.

We conclude that  
\begin{equation*}
  0=\langle \tilde z, z\rangle,
\end{equation*} and since this holds for all $\tilde z \in \mathcal D
(1,\infty)$ the proof is complete.

\end{proof}
\begin{cor}
\label{lemma:reso333bc} Suppose  $\tilde z\in
  L^2_{1+r}(1,\infty)$ where
 $r>-\tfrac{\epsilon}{2}$ with $\epsilon>0$  given
 as in \eqref{eq:q(t)}. Then there exists a uniquely determined $z\in \cup _{s<1+\tfrac{\epsilon}{2}}
 L^2_{-s}(1,\infty)$ satisfying the equation 
 \eqref{eq:z_1dim} in $\mathcal
 D'(1,\infty)$ and $z(1)=0$.
\end {cor}
\begin{proof}
  For the existence part we may assume that
  $r<\tfrac{\epsilon}{2}$. Take $z=-t^{1-r}T_r(0)t^{1+r}\tilde z$. The
  uniqueness part follows  immediately  from Lemma \ref{lemma:reso333}.
\end{proof} 
\section{Mixed problem in the case $V_2\neq 0$}
\label{Mixed problem in the case V_2 not 0} 
In the section we impose Conditions \ref{assump:conditions1} and
\ref{assump:conditions2}. We shall find an analogue of Proposition
\ref{lemma:mixed_1}. 

\subsection{Solving  a fixed point problem}
\label{Reducing to a fixed point problem} 
We are interested in solving \eqref{eq:mixed
  conditions} for $(x,\omega)\in \Gamma^+_{R,\sigma}$ where $R\geq
1$ is large and $\sigma>0$ is small. For that we write the solution
as $y=z+y_1$,
where $y_1(t)$ is the solution constructed in Section \ref{Mixed problem in the case
  V_2= V_3= 0} (with $V_2= V_3=0$). We shall derive a fixed point problem
for the ``perturbation'' $z$. By Newton's equation
\begin{align}
  \label{eq:Newtonz}
 &\ddot z =-\nabla V(z+y_1)+ \nabla V_1(y_1)=-\nabla^2 V_1(y_1)z+\mathcal R(z);\\
 &\mathcal R(z)=-\int_0^1 (1-l)\nabla^3 V_1(l z+y_1)\{z,z\}\d l-\nabla V_2(z+y_1).\nonumber
\end{align}

The Hessian in the first term on the right hand side of
\eqref{eq:Newtonz} is given
by
\begin{equation}
  \label{eq:Hessian}
 \nabla^2 V_1(y_1)=V''_1(|y_1|)P_{\|}(y_1) +|y_1|^{-1}V'_1(|y_1|)P_{\bot}(y_1), 
\end{equation} where $P_{\|}(y_1)=|y_1|^{-2}|y_1\rangle\langle y_1|$
projects onto the span of $y_1$,
and $P_{\bot}(y_1)=I-P_{\|}(y_1)$.

Using Condition \ref{assump:conditions2}, \eqref{eq:solution for r}  and the representation
\eqref{eq:Hessian} we see that $q(t):=-\nabla^2 V_1(y_1)$
satisfies the condition \eqref{eq:q(t)} with $\epsilon=\bar
\epsilon_1$. The equation \eqref{eq:Newtonz} has the form of \eqref{eq:z_1dim}
\begin{equation}
\label{eq:lin_equa}
\ddot z -qz=\tilde z :=\mathcal R(z).
\end{equation}

 We shall solve \eqref{eq:lin_equa} using Banach's  fixed point
 theorem. In this section the notation $Z_{-s}=L^2_{-s}(1,\infty)$  refers to  weighted $L^2$--spaces of $\R^d$--valued square integrable functions,
 cf. \eqref{eq:space}.

We will choose $s$ of the form 
\begin{equation}
  \label{eq:s}
 s=\alpha +\tfrac {1}{2}-\epsilon, 
\end{equation} 
where $\epsilon>0$ satisfies
\begin{align}
  \label{eq:requirement1}
 |\alpha-\tfrac{1}{2}-\epsilon|&<\tfrac{\bar\epsilon_1}{2},\\
  \label{eq:req13}
 \epsilon&< \alpha\epsilon_2. 
\end{align}

By taking  $\epsilon<\alpha\epsilon_2$
sufficiently close to $\alpha\epsilon_2$,   
indeed \eqref{eq:requirement1} and \eqref{eq:req13}   
are  fulfilled (here we use \eqref{eq:e_2small}).

We shall prove the following result.
 
\begin{prop}
   \label{lemma:mixed_2}
Suppose Conditions \ref{assump:conditions1} and 
\ref{assump:conditions2}. Fix $\epsilon>0$
sufficiently close to $\alpha\epsilon_2$ (but smaller). Then there exist $R_0\geq 1$ and $\sigma_0>0$
such that for
all $R\geq R_0 $ and for all  positive 
$\sigma\leq \sigma_0$ the problem \eqref{eq:mixed conditions} is
solved by some function $y(t)=z(t)+y_1(t),\;t\geq 1$,  for all data 
$(x,\omega)\in \Gamma^+_{R,\sigma}$ and $\lambda \geq 0$.  The
function $z(t)$ is constructed as a fixed point of
\eqref{eq:fixedpointprob} stated below. Moreover this solution 
$y(t)\in \Gamma^+_{R,\sigma}(\omega)$ for all large enough $t\geq 1$.
 \end{prop}
 \begin{proof} We shall use the operator $T_r(0)$ from Lemma
   \ref{lemma:reso4} with 
 $r=1-s$ and $s$ given by \eqref{eq:s}. Notice that then \eqref{eq:svalue} is
 fulfilled upon replacing   $\epsilon \to \bar \epsilon _1$ due to \eqref{eq:requirement1}. 

 Consider the following fixed point
problem for $z \in Z_{-s}$  
\begin{equation}
  \label{eq:fixedpointprob}
 z={\mathcal P}(z),
\end{equation} where
\begin{equation}
  \label{eq:def P} 
  {\mathcal P}(z)=-t^sT_r(0)t^{2-s} \tilde {\mathcal R}(z);\;\tilde
   {\mathcal R}(z)=\chi_1 \chi_2 {\mathcal R}(z).
\end{equation}
Here $\chi_j$  are auxiliary operators introduced in a first step; once the
fixed point is constructed they  can be removed. They are   given  in
terms of 
$z\in Z_{-s}$ by $
\chi_1=F(|z|/|y_1|<\tfrac {2}{3})$ and $
\chi_2=F(|z|/t^{\alpha-\epsilon}<2)$, respectively.

We claim that the map  ${\mathcal P}$ is a contraction on $Z_{-s}$ 
for all $(x,\omega)\in \Gamma^+_{R,\sigma}$  with $\sigma>0$ small and $R$
large, yielding by Banach's fixed point theorem a solution to
\eqref{eq:fixedpointprob}. 

We start by verifying that indeed ${\mathcal P}:
Z_{-s}\to  Z_{-s}$.

 We may
  bound the vector $\tilde {\mathcal R}(z)$ in \eqref{eq:def P}  as 
  \begin{equation}
    \label{eq:u_bound}
 \tilde {\mathcal R}(z)(t)=O\Big (t^{-\alpha(1+\mu)-2\epsilon}\Big )+O\Big (t^{-\alpha(1+\mu+\epsilon_2)}\Big ),   
  \end{equation} using  the second estimate of \eqref{eq:g/r-bound}
  and the support properties  of the $\chi_j$'s.

Since $T_r(0)$ is bounded on $L^2(1,\infty)$ we obtain from
\eqref{eq:u_bound} and \eqref{eq:req13}  that $t^sT_r(0)\* t^{2-s}  \*
\tilde
{\mathcal  R}(z)\in
Z_{-s}$. 

As for the contraction property let $z_1,z_2 \in 
Z_{-s}$  be given. Straightforward estimations  using  \eqref{eq:generalized r-bound} and \eqref{eq:req13}
 show 
\begin{equation}
  \label{eq:Lip1}
  \|{\mathcal P}(z_1)-{\mathcal P}(z_2)\|_{-s}\leq C |x|^{-\delta} \|z_1-z_2\|_{-s}\leq \tfrac{1}{2}\|z_1-z_2\|_{-s}.
\end{equation} Here we have  taken $\delta>0$ small; see 
 \eqref{eq:fixedbound1} and \eqref{eq:fixedbound2} stated below for  a similar application of
\eqref{eq:generalized r-bound}. Clearly  $C |x|^{-\delta} \leq C R^{-\delta} 
\leq  \tfrac{1}{2}$ if $R$ is large enough.

Finally we show that the factors $\chi_j$'s  in \eqref{eq:fixedpointprob} and
\eqref{eq:def P} can be removed for the constructed fixed point, say
$z=z_{-s}\in Z_{-s}$. First we notice the bound
\begin{align}\label{eq:fixedbound1}\|z\|_{-s}&\leq 2\|\mathcal P
  (z=0)\|_{-s} =2\|t^sT_r(0)t^{2-s}\nabla V_2(y_1(\cdot))\|_{-s}\nonumber\\&\leq
  C_\delta |x|^{-\delta}\leq  C_\delta R^{-\delta},
\end{align} obtained using  the contraction property \eqref{eq:Lip1}, 
  \eqref{eq:req13} and \eqref{eq:generalized r-bound}. 
We shall need a pointwise Sobolev type of bound. Let
$w(t)=\d/\d t \{t^{1-2s}|z(t)|^2\}$. By elementary  estimations and by
using \eqref{eq:fixedbound1} and Remark \ref{remark:reso5} (notice
that in conjunction with  the fixed point equation the uniform bound
of Remark \ref{remark:reso5} yields a weighted bound of the time-derivative of $z$) we may
show that 
\begin{equation*}
  \int _1^{\infty }|w(t)|dt\leq \tfrac{1}{4}\text{ for }R\text{ large enough.}
\end{equation*}

From this estimate we get (by integrating to infinity)
\begin{equation}
  \label{eq:lest}
  |z(t)|\leq \tfrac{1}{2}t^{\alpha-\epsilon};\;t\geq 1.
\end{equation}

Combining \eqref{eq:lest} with the bound
\begin{equation}\label{eq:fixedbound2}
  t^{\epsilon-\alpha}|y_1|\geq c|x|^{\epsilon/\alpha}\geq
  cR^{\epsilon/\alpha}\geq 2,
\end{equation} we conclude  that indeed $
  \chi_1=F(|z|/|y_1|<\tfrac{2}{3})=1$ and
  $\chi_2=F(|z|/t^{\alpha-\epsilon}<2)=1$ for all 
 sufficiently large $R$'s. Consequently those factors $\chi_j$'s
  can be removed.

Obviously $z(1)=0$  and the problem \eqref{eq:mixed conditions} is
solved by $y(t)=z(t)+y_1(t)$. 
 \end{proof}

\begin{remarks}\label{remarks:remark_fixed}
\begin{enumerate}[\quad\normalfont 1)]
\item \label{it:delta_bound} The above analysis yields the following uniform bound of the fixed point (with $\epsilon$ as above)
\begin{equation*}|z(t)| \leq C_\delta |x|^{-\delta}t^{\alpha-\epsilon},
\end{equation*}  valid for some
$\delta=\delta(\epsilon)>0$.
\item \label{it:delta_bound111} For 
   positive energies there is a simpler procedure, cf. \cite [proof of Theorem 1.5.1]{DG}. This leads to the improved
  decay in time
  \begin{equation}
    \label{eq:posimprv}
    y(t)-(t-1)\sqrt{2\lambda}\omega-x=O(t^\delta);\;\delta>\max(1-\mu,0),
  \end{equation}
 with the  bounding constant being locally
  uniform in $(\omega,\lambda)\in S^{d-1}\times (0,\infty)$. Obviously
  (\ref{eq:posimprv}) is not uniform in $\lambda$. Compared to the
  procedure for positive energies the present one is based
  on an additional Taylor expansion. In this way we circumvent  a problem
  related to the fact that the quantity $\int t|\nabla ^2V(y)|dt$ is finite only
  for $\lambda>0$ (causing a difficulty for the contraction property at $\lambda=0$).
  
\item \label{it:invariance}Although it is not stated in Proposition  \ref{lemma:mixed_2} that
  $\Gamma^+_{R,\sigma}(\omega)$ is invariant under the forward flow  this is
  indeed true; see Lemma \ref{lemma:mixed_22} stated below. Notice that it follows from  Proposition 
  \ref{lemma:mixed_1} that 
  $\Gamma^+_{R,\sigma}(\omega)$ is invariant in the case $V_2=0$. 
\item \label{it:non-uniqueness} We have not proved that the solution
  to  the problem \eqref{eq:mixed conditions} is unique in the sense
  used in Proposition  \ref{lemma:mixed_1} in the case
  $V_2=0$. 
\end{enumerate}
\end{remarks}

\begin{defn} \label{def:vector field}Under the conditions of Proposition   \ref{lemma:mixed_2}
  we define a vector field $F$ on $\Gamma^+_{R_0,\sigma_0}(\omega)$ by
  \begin{equation}
    \label{eq:vector field}
    F(x)=\dot y(t=1;x,\omega,\lambda);
  \end{equation} here $y$ refers to the solution of \eqref{eq:mixed
    conditions} given in Proposition \ref{lemma:mixed_2}.
\end{defn}
\begin{lemma}
\label{lemma:mixed_22} Let 
  $y=y(t)=y(t;x,\omega,\lambda)$ be  the  solution of 
  Proposition  \ref{lemma:mixed_2}. Then $y \in \Gamma^+_{R,\sigma}(\omega)$
  for all  $t\geq 1$. 

Let $F_1$ be given as in Definition
\ref{def:vector field} in the case  $V_2=0$, and let $\epsilon$ be
given as in Proposition  \ref{lemma:mixed_2}. Then for all positive
$\epsilon'<\epsilon$ and $\epsilon_2'<\epsilon_2$ 
\begin{equation}
  \label{eq:F's}
 F(x)-F_1(x)=O\big(|x|^{-\mu/2-\breve \epsilon }
\big);\;\breve \epsilon := \min (\epsilon'/\alpha,\;\epsilon'_2).
\end{equation} In particular for constants $C,\;c>0$ independent of $x,\;\omega
$ and $\lambda$
\begin{equation}
  \label{eq:F's2}
  \Big|\frac {F(x)}{|F(x)|}-\frac {F_1(x)}{|F_1(x)|}\Big |\leq
  C|x|^{-\breve \epsilon},
\end{equation} and 
\begin{subequations}\label{group2}
\begin{align}
\label{eq:F_angle1}
 \frac {F(x)}{|F(x)|}\cdot \hat x&\geq 1-C\big (1-\hat x \cdot \omega \big )- C|x|^{-\breve \epsilon},\\ 
\label{eq:F_angle2}
   \frac {F(x)}{|F(x)|}\cdot \hat x  &\leq 1-c\big (1-\hat x \cdot \omega \big )+ C|x|^{-\breve \epsilon},\\ 
\label{eq:F_angle3}
 \frac {F(x)}{|F(x)|}\cdot \omega &\geq 1-C\big (1-\hat x \cdot \omega \big )- C|x|^{-\breve \epsilon}. 
\end{align} 
\end{subequations}
\end{lemma}
\begin{proof} Let $y_1=y_1(t)$ signify the solution in the case
  $V_2=0$. From \eqref{eq:F_rep} and Taylor's formula we obtain 
  \begin{equation}
    \label{eq:y_dot'ss}
    \dot y(t) -\dot y_1(t)=\int _t^\infty \Big \{\int _0^1\nabla
    ^2V_1(lz+y_1)z \d l+\nabla V_2(y)
\Big \}\,\d s.  \end{equation}

To bound the contribution from the first term on the right hand side
we use 
(\ref{eq:g/r-bound}) and \eqref{eq:lest}, and estimate with  $\delta=
  2^{-1}(1-\alpha+\epsilon')$ for $\epsilon'<\epsilon$ 
\begin{align}
\label{eq:nabla^2V_1}
\int _t^\infty \int _0^1\nabla
    ^2V_1(lz+y_1)z \,\d l\d s
  &=
\int _t^\infty O\big (|y_1|^{-\delta(2+\mu)}\big
  )s^{-(1-\delta)\alpha(2+\mu)}s^{\alpha-\epsilon}\d s\nonumber 
\\
&=O\big
(|y_1(t)|^{-\mu/2-\epsilon'/\alpha}\big ).
\end {align}

The contribution from the second  term on the right hand side is
estimated similarly
\begin{align}\label{eq:nablaV_2}
\int _t^\infty \nabla V_2(y)\,\d s  
  &=|y_1(t)|^{-\mu/2-\epsilon'_2}\int _t^\infty O\big
  (|y_1|^{-1-\mu/2 +\epsilon'_2-\epsilon_2}\big
  )\,\d s \nonumber 
\\&=O\big (|y_1(t)|^{-\mu/2-\epsilon'_2}).
\end {align}

We conclude \begin{equation}
    \label{eq:y_dot'ss2}
    \dot y(t) -\dot y_1(t)=O\big
(|y_1(t)|^{-\mu/2-\breve \epsilon)}\big ) =|\dot y_1(t)|O\big(|x|^{-\breve \epsilon}
\big).  \end{equation}

We obtain \eqref{eq:F's} by taking $t=1$ in
\eqref{eq:y_dot'ss2}. Clearly \eqref{eq:F's2} follows from
\eqref{eq:F's}. Moreover  \eqref{eq:F_angle1} and \eqref{eq:F_angle2} in turn follow from
\eqref{eq:F's2} and Section \ref{Mixed problem in the case V_2= V_3=
  0} 
(possibly after diminishing $\sigma_0$), while \eqref{eq:F_angle3} readily follows
from \eqref{eq:F_angle1} (for a new constant). Notice for
\eqref{eq:F_angle1} and \eqref{eq:F_angle2} in the case $V_2=0$ that
$1-\frac {F(x)}{|F(x)|}\cdot \hat x  =1-\cos \psi_1$ and
$1-\hat x \cdot \omega =1-\cos \theta_1$. Whence the statements are equivalent
to the bounds $c\theta_1\leq \psi_1\leq C\theta_1$ which may be
derived from the following formula (representing $\kappa=-\sin \psi_1$)
\begin{equation}
  \label{eq:kap-the0}{\partial \kappa^2\over \partial
    \theta_1^2}(\theta_1=0)=\Big( \int_1^\infty s^{-2}{g(r_1)\over
    g(sr_1)}\;\d s\Big)^{-2}.
  \end{equation}

 Finally we obtain  from \eqref{eq:F's2}, \eqref{eq:y_dot'ss2}  and the above 
  considerations (for the case $V_2=0$) that $y(t) \in \Gamma^+_{R,\sigma}(\omega)$
  for all  $t\geq 1$ given that $x=y(1) \in \Gamma^+_{R,\sigma}(\omega)$. 

\end{proof}
We shall show in Section \ref{Solution to the eikonal equation} that $F$ is a smooth gradient field. The following
result, the proof of which is somewhat complicated since we have
not proved uniqueness, cf. Remarks \ref {remarks:remark_fixed} \ref
{it:non-uniqueness}), will be useful.

\begin{lemma} \label{lemma:mixed_3} 
Let 
  $y=y(t)=y(t;x,\omega,\lambda)$ be  the  solution of 
  Proposition  \ref{lemma:mixed_2}. Then $\dot y(t)=F(y(t))$
  for all $t\geq 1$.
\end{lemma}
\begin{proof} Let us omit $\omega,\lambda$ in the notation, and 
  consider the following equivalent statement, say $p(T)$,
  \begin{equation}
    \label{eq:group+prop}
    y(t+\bar t-1;x)=y(t;y(\bar t;x))\;\text {for all}\;t\geq1\;\text {and all}\;\bar t\in [1,T].
  \end{equation} Here $T\geq 1$ is arbitrary. 

Obviously $p(1)$ is true. Let
  us prove that $p(T)$ is true for a $T>1$  that may be chosen to be  independent of $x$:
We consider 
\begin{equation*}
 \tilde z
(\cdot):=y(\cdot+\bar t-1;x)-y_1(\cdot;y(\bar t;x)) 
\end{equation*}
 for $\bar t\in (1,T]$. We claim
that (with $s$ given by \eqref{eq:s})
\begin{align}
  \label{eq:tilde zin Z_1}
  \tilde z&\in Z_{-s},\\
\label{eq:tilde zin Z_2}
|\tilde z|&<\tfrac {1}{3}|y_1(\cdot;y(\bar t;x))|,\\
\label{eq:tilde zin Z_222}
|\tilde z|&<t^{\alpha-\epsilon}.
\end{align}

 Notice that by using \eqref{eq:tilde zin
 Z_1}--\eqref{eq:tilde zin Z_222}, the fact that  $\tilde z(1)=0$,
 Lemma \ref{lemma:reso333} and the uniqueness property for 
contractions  we obtain that $\tilde z(t)=z(t;y(\bar t;x))$ and therefore
 indeed \eqref{eq:group+prop} (for suitably small $T-1>0$). Here Lemma
 \ref{lemma:reso333}  is applied to the vector $\tilde z- \mathcal
 P (\tilde z)$.

We estimate 
\begin{align}
  \label{eq:tilde zin Z_3}
  |\tilde z(t)|\leq{}& |y(t+\bar t-1;x)-y(t;x)|\\
&+|z(t;x)|+|y_1(t;x)-y_1(t;y(\bar t;x))|,\nonumber
\\
  \label{eq:dif_y1}
  |y(t+\bar t-1;x)-y(t;x)|\leq {}&
  \int_0^{\bar t-1} |\dot y(s+t;x)|\d s=O(t^0),
\end{align}
and
\begin{align}
  \label{eq:dif_y2}
  y_1(t;x)-y_1(t;y(\bar t;x))&=\int_0^{1} (\nabla_x
  y_1)(t;l(x-y(\bar t;x))+y(\bar t;x))\cdot (x-y(\bar t;x))\d l
  \nonumber\\
  &=O\left(g(|y_1|)^{-1}\right)=O(t^{\alpha\mu/2}),
\end{align}
cf. \eqref{eq:y_x_der}. 

We obtain \eqref{eq:tilde zin Z_1} from \eqref{eq:tilde zin Z_3}--\eqref{eq:dif_y2}. 

As for  \eqref{eq:tilde zin Z_2} we may use the estimates
\begin{align}
  \label{eq:dif_y3}
  |y_1(t;y(\bar t;x))|&\geq |y_1(t;x)|-|y_1(t;y(\bar t;x))-y_1(t;x)|,
  \\
  \label{eq:dif_y4}
  |z(t;x)|&\leq \tfrac {1}{4}|y_1(t;x)|,
\end{align}
and the previous estimates. (Here the smallness of
$T-1>0$ comes in.) The proof of \eqref{eq:tilde zin Z_222} is similar.

Now to show \eqref{eq:group+prop} in the general case, suppose $p(T)$
for some $T>1$:
Then for $\triangle \bar t>0$ small (in agreement with the
previous step) we have, with $\bar t=T+\triangle \bar t$,
\begin{align*}
 y(t+\bar t-1;x) &=y((t+\triangle \bar t)+T-1;x)\\
& =y(t+\triangle \bar t;y(T;x))=y(t;y(\triangle \bar t+1;y(T;x)))\\
&=y(t;y(\bar t;x)).
\end{align*} Here we used $p(T)$ as well as the previous step with $x$
replaced by $y(T;x)$. Whence we have shown $p(T')$ for a $T'>T$, and
therefore \eqref{eq:group+prop} for all $T\geq 1$.
\end{proof}
\subsection{Smoothness properties of solution $y$}
\label{Smoothness properties of solution $y$}
We shall compute and estimate derivatives with respect to
initial position $x$ and final direction $\omega$ of  the constructed
solution $y=z+y_1$ and of the vector field $F$ given in Definition
\ref{def:vector field}. We
studied the derivatives of $y=y_1$ in Subsection \ref{Dependence of flow
  on data}. It is well-known that under general conditions a solution
to a fixed point equation depending on parameters will be smooth in
these variables, see for instance \cite[Appendix C]{I}.

From the fixed point equation 
\eqref{eq:fixedpointprob}  
one may derive (for example)  the
representation
\begin{equation}
  \label{eq:formu_der}
\partial _x z=(I-\nabla _z\mathcal P)^{-1} \partial _x\mathcal P.  
\end{equation}
Notice here the bound
\begin{equation}\label{eq:zderformu_der}
\|\nabla _z\mathcal P\|_{\mathcal {B}(Z_{-s})}\leq \tfrac{1}{2},  
\end{equation} cf.  \eqref{eq:Lip1} (with $s$ given by
\eqref{eq:s}). To deal with higher order  derivatives we need a more
elaborate analysis.

Motivated by \eqref{eq:lest} we introduce the following 
modification of the spaces $Z_{-\sigma}$ of \eqref{eq:space}. Let $
Z_{-\sigma}^{\unif}$ be the space of   $\R^d$--valued continuous
functions $\tilde z$ on
$[1,\infty)$ obeying $$\|\tilde z\|_{-\sigma}^{\unif}:=\sup_{t\geq
  1}t^{-\sigma}|\tilde z(t)|<\infty.$$ We shall first estimate various
derivatives of the contraction $\mathcal P$ on $Z_{-s}$.
\begin{lemma}
  \label{lemma:derofz1} For all multiindices
  $\delta$ and $\gamma$, $k\in \N\cup\{0\}$   and $
  z_1,\dots ,
  z_k\in 
Z_{\epsilon-\alpha}^{\unif}\cap Z_{-s}$
\begin{align}
  \label{eq:der_z_x} \|\partial _\omega^\delta
      \partial _x^\gamma \partial_z^k \mathcal
  P\{
  z_1,\dots ,
  z_k\}\|_{-s}&\leq C_{\delta,\gamma,k} |x|^{-|\gamma|}\|
    z_1\|_{\epsilon-\alpha}^{\unif}\cdots \|
    z_k\|_{\epsilon-\alpha}^{\unif},\\
\label{eq:der_z_x111} 
\|\tfrac {\d }{\d t}\partial _\omega^\delta
      \partial _x^\gamma \partial_z^k \mathcal
  P\{
  z_1,\dots ,
  z_k\}\|_{1-s}&\leq C_{\delta,\gamma,k} |x|^{-|\gamma|}\|
    z_1\|_{\epsilon-\alpha}^{\unif}\cdots \|
    z_k\|_{\epsilon-\alpha}^{\unif}.
\end{align}
\end{lemma}
\begin{proof}
 We start by verifying \eqref{eq:der_z_x} for $k=0$, $\delta=0$ and
 $|\gamma|=1$.  So we need to trace the $x$--dependence of
 $\mathcal P$ as defined by \eqref{eq:def P}. There is a contribution
 from differentiating the factor $T_r(0)$ and another from
 differentiating the factor $\tilde {\mathcal R}(z)$. Using Lemma
 \ref{lemma:reso4} we may use the formal computation
 \begin{equation}
   \label{eq:formcoT}
  \partial _xT_r(0)=-T_r(0)t (\partial _x q)tT_r(0);
 \end{equation} here 
 \begin{equation}
   \label{eq:x q}
   \partial _x q=-\nabla^3 V_1(y_1) \partial _x y_1.
 \end{equation} 

Using \eqref{eq:g/r-bound}, \eqref{eq:y_x_derhigher} and \eqref{eq:x q} we derive $t
(\partial _x q)t=O(|x|^{-1})$. Whence $\partial
_xT_r(0)=O(|x|^{-1})$. As for the $x$--dependence from the factor
$\tilde {\mathcal R}(z)$ we may combine \eqref{eq:y_x_derhigher} and
the arguments for \eqref{eq:u_bound} to pick up an extra factor
$|x|^{-1}$ in the estimation of  $\partial _x\tilde {\mathcal R}(z)$.

Higher derivatives are treated similarly. 

As for \eqref{eq:der_z_x111}
we use Remark \ref{remark:reso5} and the same estimates as before.
 
\end{proof}
\begin{lemma}
  \label{lemma:derofz2} For all multiindices
  $\delta$ and $\gamma$
\begin{subequations}\label{group}
\begin{align}
\label{eq:der_z_xhigher1} \|\partial _\omega^\delta  \partial _x^\gamma
  z\|_{-s}&= O\left(|x|^{-|\gamma|}
   \right ),\\
\label{eq:der_z_xhigher2} \|\partial_t\partial _\omega^\delta  \partial _x^\gamma
  z\|_{1-s}&= O\left(|x|^{-|\gamma|}
   \right ),\\
\label{eq:der_z_xhigher} \|\partial _\omega^\delta  \partial _x^\gamma
  z\|_{\epsilon-\alpha}^{\unif}&= O\left(|x|^{-|\gamma|}
   \right ).
\end{align}
\end{subequations}
\end{lemma}
\begin{proof} We notice that
   \eqref{eq:der_z_xhigher1}--\eqref{eq:der_z_xhigher} in the case
  $|\delta|+|\gamma|=0$ follow from \eqref{eq:fixedbound1},
  \eqref{eq:lest} and the arguments for \eqref{eq:lest}. 

By the same
  reasoning (the Sobolev bound) if \eqref{eq:der_z_xhigher1} and \eqref{eq:der_z_xhigher2}
  are known for $|\delta|+|\gamma|\leq n$ for some
  $n\in 
\N$, then also  \eqref{eq:der_z_xhigher} for $|\delta|+|\gamma|\leq n$
  is valid.

  So suppose we know \eqref{eq:der_z_xhigher1}--\eqref{eq:der_z_xhigher}
  for all multiindices
  $\delta$ and $\gamma$ with $|\delta|+|\gamma|\leq n-1$ for some
  $n\in 
\N$, then we only need
  to verify the bounds \eqref{eq:der_z_xhigher1} and
  \eqref{eq:der_z_xhigher2} for $|\delta|+|\gamma|\leq n$. For this we
  fix multiindices $\delta$ and $\gamma$ with $|\delta|+|\gamma|= n-1$
  and look at the representation of $\breve z=\partial _\omega^\delta  \partial _x^\gamma
  z$ obtained from differentiating \eqref{eq:fixedpointprob} (a Faa di
  Bruno formula)
  \begin{align}
    \label{eq:faadibruno}
    \breve z= {} &(\partial_z\mathcal P)\{\breve z\}+\partial
    _\omega^\delta  \partial _x^\gamma \mathcal P\\&+\sum
    c_{\delta',\delta_1,\dots,\delta_k,\gamma',\gamma_1,\dots, \gamma_k}(\partial
    _\omega^{\delta'}  \partial _x^{\gamma'} \partial_z^k \mathcal P)\{\partial _\omega^{\delta_1}  \partial _x^{\gamma_1}
  z,\dots,\partial _\omega^{\delta_k}  \partial _x^{\gamma_k}
  z\},\nonumber
  \end{align} where summation is over $k\geq1,
    \delta'+\delta_1+\cdots +\delta_k=\delta,\gamma'+\gamma_1+\cdots
    +\gamma_k=\gamma$ and $ n-1\geq k+|\delta'|+|\gamma'|\geq 2$. The
    meaning of \eqref{eq:faadibruno} if $n=1$ is
    \eqref{eq:fixedpointprob}, while for  $n=2$ the third term to the
    right should be omitted. Now
    we may compute $\partial \breve z$ (meaning either $\partial_\omega^{e_i}
    \breve z$ or $\partial_x^{e_j} \breve z$) by differentiating
    \eqref{eq:faadibruno}. The result is, cf. \eqref{eq:formu_der}, 
    \begin{equation*}
      \partial \breve z=(\partial_z\mathcal P)\{\partial \breve
      z\}+\tilde z,
    \end{equation*} where $\tilde z$ may  be treated using
      \eqref{eq:der_z_x} and the induction hypothesis. So (again) we
      may invoke \eqref{eq:zderformu_der}. This yields
      \eqref{eq:der_z_xhigher1} (as well as the representation
      \eqref{eq:faadibruno}) for $|\delta|+|\gamma|= n$. 

It remains to prove \eqref{eq:der_z_xhigher2} in the inductive
argument. For that we use the proven formula  \eqref{eq:faadibruno}
for $|\delta|+|\gamma|= n$. We proceed somewhat similarly applying now
$t\partial_t$ to both sides of this formula with $\breve z$ now given
in terms of indices  with $|\delta|+|\gamma|= n$.
 This leads to 
\begin{equation*}
      t\partial_t \breve z=t\tfrac {\d }{\d t} \partial_z\mathcal P\{ \breve
      z\}+t\partial_t \bar z.
    \end{equation*} The first term to the right may  be treated using
      Remark \ref{remark:reso5}; it is estimated as 
      \begin{equation*}
        \|t\tfrac {\d }{\d t}\partial_z\mathcal P\{ \breve
      z\}\|_{-s}\leq C\|\breve
      z\|_{-s},
      \end{equation*} cf. \eqref{eq:Lip1}. The second term may  be
      treated using  \eqref{eq:der_z_x111} and the induction
      hypothesis (specifically only \eqref{eq:der_z_xhigher}). The
      estimate \eqref{eq:der_z_xhigher2} follows.
\end{proof}
\begin{prop}  
  \label{prop:derofz}
With $F$  being the vector field  
in  Definition
\ref{def:vector field} there are uniform bounds valid for all multiindices $\delta$ and $\gamma$ 
 \begin{align}\label{eq:derF222}\partial _\omega^\delta \partial _x^\gamma  F(x)&=\langle x \rangle ^{-|\gamma|}O\left(
   g(|x|) \right ),\\
\label{eq:derF2222} \partial _\omega^\delta \partial _x^\gamma  \big (F(x)-F_1(x)\big
   )&=\langle x \rangle ^{-\breve \epsilon-|\gamma|}O\left(
   g(|x|) \right );
\end{align}  
 here $F_1$ is given as $F$ for the case $V_2=0$, and
$\breve \epsilon>0$ is given as in Lemma \ref{lemma:mixed_22}. 
\end{prop}           
\begin{proof} 
  As for \eqref{eq:derF222} we shall use the same scheme as for
  proving 
\eqref{eq:y_x_derhigher2}. First we notice the following consequence of
  \eqref{eq:der_z_xhigher}.
  \begin{equation}
    \label{eq:y1bnd1}
|\partial _\omega^\delta  \partial _x^\gamma
  z(t)|\leq C_{\delta,\gamma}|y_1(t)|\;|x|^{-|\gamma|}.
  \end{equation}

By \eqref{eq:y_x_derhigher}
\begin{equation}
    \label{eq:y1bnd2}
|\partial _\omega^\delta  \partial _x^\gamma
  y_1(t)|\leq C_{\delta,\gamma}|y_1(t)|\;|x|^{-|\gamma|}.
  \end{equation}

The combination of \eqref{eq:y1bnd1} and \eqref{eq:y1bnd2} is
\begin{equation}
    \label{eq:y1bnd3}
|\partial _\omega^\delta  \partial _x^\gamma
  y(t)|\leq C_{\delta,\gamma}|y_1(t)|\;|x|^{-|\gamma|}.
  \end{equation}

As in the proof of
\eqref{eq:y_x_derhigher2},
we represent 
\begin{equation}
  \label{eq:der_xF_rep23}
  \partial_{*} F=\partial_{*} \dot y(t=1)=\partial_{*} \sqrt{2\lambda}
  \omega +\int _1^\infty \partial \nabla
  V(y)\partial_{*}  y \,\d t,
\end{equation} from which we may derive a Faa di Bruno formula (by
  repeated differentiation) to
  which \eqref{eq:y1bnd3} applies. The argument for  the
  case $\delta=0$ and $|\gamma|=1$ is similar to \eqref{eq:dot y_x_der}:

By combining \eqref{eq:y1bnd3} and \eqref{eq:der_xF_rep23} we obtain 
\begin{align}\label{eq:derdot_y_111}
  \partial _x \dot y(t=1;x,\omega,\lambda)&=\int _1^\infty \nabla^2
  V(y)O\Big (\frac{|y|}{|x|}\Big )\d t\nonumber \\
&=O\Big (|x|^{-\big (1+\tfrac {\mu}{2}\big )}\Big )=\langle x\rangle ^{-1}O(g(|x|)),
\end{align} which obviously is  a particular case of
  \eqref{eq:derF222}. The general case is similar.  

As for \eqref{eq:derF2222} we need a more refined argument than 
\eqref{eq:derdot_y_111}; this is now  
based on  \eqref{eq:y_dot'ss}.  We need to differentiate and
estimate the expressions to the left in \eqref{eq:nabla^2V_1} and
\eqref{eq:nablaV_2}. The estimation of the differentiated expressions is done  by using \eqref{eq:der_z_xhigher} and
\eqref{eq:y1bnd2}  in a
similar manner as done in the proof of Lemma
\ref{lemma:mixed_22}; details are omitted.  
\end{proof} 

\begin{lemma}
 \label{lemma:appron2}  The vector field
 $F=F(x,\omega,\lambda)$ as well as all derivatives $\partial
 _\omega^\delta \partial _x^\gamma F$ are jointly continuous in the  variables $(x,\omega)\in \Gamma^+_{R_0,\sigma_0}$ and $\lambda \geq 0$.
\end{lemma}

\begin{proof} Since in fact $F(x,\omega,\lambda)$ is smooth in
 $(x,\omega)\in \Gamma^+_{R_0,\sigma_0}$ and $\lambda > 0$, cf. Remarks
 \ref{remark:nonsmooth} \ref{it:konkap1}),  only
 continuity at $\lambda=0$ is non-trivial. Due to
 Remarks \ref{remark:nonsmooth} \ref{it:konkap2}) and Proposition
 \ref{prop:derofz} it suffices to show that 
 \begin{equation}
   \label{eq:Funfif2}
  F(x,\omega,\lambda)\to F(x,\omega,0)\text{ for }\lambda \to 0. 
 \end{equation}

For that we first notice that 
\begin{equation}
   \label{eq:zccco2}
 y_1(t;x,\omega,\lambda)\to
y_1(t;x,\omega,0)\text{ for }\lambda \to 0.
\end{equation} This may be seen by
combining Remarks
 \ref{remark:nonsmooth} \ref{it:konkap1}) and a standard continuity
 statement of
  a flow  in terms of variation of the initial values  and the
  vector field, see for example \cite[Theorem 3 page 177]{BR}. 

Since $\mathcal P=\mathcal P(\zeta,\lambda)\in Z_{-s}$ is jointly continuous in
 $\lambda \geq 0$ and $\zeta\in Z_{-s}$ (as may readily be checked)
 and there is a uniform contraction constant, a general principle for contractions, cf. \cite[Appendix
 C]{I}, yields continuity for the fixed points; viz.  
 \begin{equation}
   \label{eq:zccco}
 z_\lambda\to z_0=z_{\lambda=0} \text{ in } Z_{-s}.  
 \end{equation}

Next we represent, cf. \eqref{eq:y_dot'ss},
\begin{equation*}
    F(x,\omega,\lambda) -F(x,\omega,0)=\int _1^\infty \big (\nabla V(y_\lambda)-\nabla V(y_0)
\big )\d t. 
\end{equation*}
The norm of the integrand to the right is estimated uniformly by
$Ct^{-\alpha(1+\mu)}$. Combining this fact, \eqref{eq:zccco2},
\eqref{eq:zccco} and \cite[Theorems 1.34, 3.12]{R} we conclude \eqref{eq:Funfif2}.   
\end{proof}

We end this section by stating a somewhat similar approximation result
needed in the next section; clearly there are  results for
higher derivatives as in Lemma \ref{lemma:appron2} but they will not
be needed.
Let $V_{2,n}(x)=F(|x|/n<1)V_2(x)$ for   $n \in
\N$, and let $z_n, y_n, \mathcal P_n$ and $F_n$ be the quantities defined upon
replacing $V_2$ by  $V_{2,n}$ in previous constructions.
\begin{lemma}
 \label{lemma:appron}  The vector field
 $F_n=F_n(x,\omega,\lambda)$ is defined on the same domain as $F$
 (possibly after a slight shrinking),
 and pointwisely $$\partial_x F_n\to \partial_x F\text { for }n\to\infty.$$
\end{lemma}
\begin{proof} Clearly for all multiindices $\gamma$, the function
  $\langle x\rangle ^{\mu+\epsilon_2+|\gamma|}\partial^\gamma
  V_{2,n}(x)$ is bounded uniformly in $n$. Using this property one may
  check the first statement as well as the existence of uniform bounds
  on the quantities $\sup_x \langle x\rangle ^{|\gamma|}\|\partial_x
  ^\gamma z_n\|_{-s}$ and $\sup_x g(x)^{-1}\*\langle x\rangle
  ^{|\gamma|}\*|\partial^\gamma F_n(x)|$. Whence it suffices to show
  that
 \begin{equation}
   \label{eq:Funfif}
  F_n(x)\to F(x)\text{ for }n\to\infty, 
 \end{equation} cf. Remarks \ref{remark:nonsmooth} \ref{it:konkap2}).

Since $\mathcal P_n(\zeta)\in Z_{-s}$ is jointly continuous in
 $n\in \N$ and $\zeta\in Z_{-s}$ (more precisely $\|\mathcal P_n(\zeta_n)-
 \mathcal P(\zeta)\|_{-s}\to 0$ for any sequence $\zeta_n\to \zeta$ in
 $Z_{-s}$) we have continuity for the fixed points; viz.  $z_n\to z$ in $Z_{-s}$.

We represent
\begin{equation*}
    F_n -F=\int _1^\infty \big (\nabla V_n(y_n)-\nabla V(y)
\big )\d t. 
\end{equation*} As in the proof of Lemma \ref{lemma:appron2} we have a
uniform bound and pointwise convergence (along  subsequences) for the integrand; we can
argue as before and conclude \eqref{eq:Funfif}.   


%
\end{proof}

\section{Solution to  eikonal equation}
\label{Solution to the eikonal equation}

In this section we shall see that the vector field $F$ of Definition
\ref{def:vector field} can be written as $F(x)=\nabla \phi(x)$ for some smooth
function  $\phi$. We impose
Conditions \ref{assump:conditions1} and \ref{assump:conditions2},
however if $V_2 =0$  Condition \ref{assump:conditions1} suffices; in
that  case $F$ is given by the same  definition.

\begin{defn}
  \label{defn:phasefunc}
Under the conditions of Proposition 
  \ref{lemma:mixed_2} (or  Proposition \ref{lemma:mixed_1}) we introduce
           for
  $(x,\omega)\in \Gamma^+_{R_0,\sigma_0}$ and $\lambda\geq 0$
  \begin{equation*}
    \phi(x)=\phi(x,\omega,\lambda)=(x-R_0\omega)\cdot\int _0^1
    F(l(x-R_0\omega)+R_0\omega)\d l +\sqrt {2\lambda}R_0.
  \end{equation*} 
\end{defn}

It follows from Lemma \ref{lemma:appron2} that $\phi=\phi(x,\omega,\lambda)$ as well as all derivatives $\partial
 _\omega^\delta \partial _x^\gamma \phi$ are jointly continuous in the 
 variables $(x,\omega)\in \Gamma^+_{R_0,\sigma_0}$ and $\lambda \geq
 0$. We shall show that the image of the map
 $\Gamma_{R_0,\sigma_0}^+(\omega)\ni x\to (x,F(x)$ is Lagrangian, so that
 indeed this function $\phi$ is an antiderivative  of  $F$.

\begin{prop} \label{prop:phase}Under the conditions of Definition
    \ref{defn:phasefunc} $$F(x)=\nabla_x \phi(x),$$  and $\phi$ solves the
    eikonal equation
    \begin{equation}
      \label{eq:eikonal}
  \tfrac{1}{2}(\nabla_x \phi)^2 +V(x)=\lambda;\;x \in \Gamma^+_{R_0,\sigma_0}(\omega).
    \end{equation}
\end{prop}
\begin{proof}
 Let us denote by $\theta_t=(y,F(y))$ the Hamiltonian orbit located 
at  time $t=1$ at the point $(x,F(x))$, cf. Lemma \ref {lemma:mixed_3}. Viewing
$\theta_t=\theta_t(x)$ as a function of  $x$ we shall  show that 
\begin{equation}
  \label{eq:symplec1}
  \theta_{1}^*\sigma=0,
\end{equation} where here $\sigma=\sum d\xi_i\wedge dx_i$ is the canonical
two-form. For that we invoke the  continuity property in the dependence
through the term $V_2$ as specified in Lemma \ref{lemma:appron}. 
We obtain that $\theta_{1}^*\sigma = \lim_{n\to \infty} \theta_{1,n}^*\sigma$ (using obvious
notation), and  henceforth 
we may assume that $V_2$ is compactly supported.

Next, since
$\theta_{1}^*\sigma=\theta_{t}^*\sigma$ for all $t\geq 1$ it 
suffices to show that the strong limit
\begin{equation}
  \label{eq:symplec2}
 \lim_{t\to \infty}\theta_{t}^*\sigma =0.
\end{equation}
We pick $\bar t >1$  so large that the first coordinate, say $\bar x$,
of $\theta_{\bar t}(x)$ is outside the support of $V_2$ (and similarly
for all later times). Considering $\bar x=\bar x(x)$ as a function of
$x$ we may write $\theta_{t}^*\sigma =\bar x^*\theta_{t-\bar t+1}(\bar x)^*\sigma$, cf. \eqref{eq:group+prop},  and compute 
\begin{equation*}
 \theta_{t-\bar t+1}(\bar x)^*\sigma
 =\sum_{k<l}\partial_{\bar x_l}y\cdot(F'-F'^{\tr})\partial_{\bar
 x_k}y\;\;\d \bar x_k\wedge \d \bar x_l.
\end{equation*} Here $F'$ signifies the derivative of $F$ at $y$, and
``tr'' is used for the transposed operator.
Now, using \eqref{eq:y_x_der} and \eqref{eq:derF222} we get
\begin{equation*}
 \partial_{\bar x_l}y\cdot(F'-F'^{\tr})\partial_{\bar x_k}y=O(g(|y|)|y|^{-1})O\left(\frac{g(|\bar x|)^2}{g(|y|)^2
}\right)= O\left(\frac{g(|\bar x|)^2}{|y|g(|y|)
}\right).
\end{equation*} The right hand side $\to 0$, and
therefore \eqref{eq:symplec2} follows. 
\end{proof}


\begin{remarks}\label{remarks:remark_eikonal}
\begin{enumerate}[\quad\normalfont 1)]
\item \label{it:delta_bound2}For $\lambda>0$ the constructed phase function 
  essentially coincides with the Isozaki Kitada (outgoing) phase function, $\phi(x,\xi),\;\xi=\sqrt{2\lambda}\omega$,
  cf. \cite[Definition 2.3]{IK1} or \cite[Proposition 2.8.2]{DG}. In
  particular, cf. the method of proof of 
Proposition \ref{prop:derofz},  there are bounds 
  \begin{align}\label{eq:postca}
    \partial_\lambda^k\partial_\omega^\kappa \partial_x^\gamma\{\phi(x,\omega,\lambda) -\sqrt{2\lambda}x\cdot
    \omega\}&=O\big(|x|^{\delta-|\gamma|}\big )\;\text {for}\;|x|\to
    \infty;\\\delta&>\max (1-\mu,0).\nonumber   
  \end{align}
However these bounds are not  uniform in  $\lambda$ as opposed 
to \eqref{eq:derF222}. (We paid in
\eqref{eq:derF222}  the price of weaker pointwise decay.) 
\item \label{it:alternativeLagrange}We constructed  the phase by
  integrating the vector field $F$. In \cite{IK1} and  \cite{DG} it is
  constructed by a different procedure. Since it is assumed there that
  $\lambda$ keeps away from $0$ one would need additional elaboration
  to include $\lambda=0$ by that procedure. Our arguments are 
  related to \cite[p. 16]{Hel} and \cite[proof of Theorem 2.1]{HS}.
\item \label{it:phiasympt} We may 
      integrate from $R_0\hat x$ to $x$ along the line segment joining
      the two points plus in addition on the (small) arc
      joining $R_0\omega$ and $R_0\hat x$ on a great circle of radius
      $R_0$. This gives the following represention  in the case $V_2=0$ 
\begin{align} \label{eq:phirepresen}
       \phi(x,\omega,\lambda)&=\tilde\phi(r,\hat x\cdot
       \omega,\lambda) +\phi_2(\hat
       x,\omega,\lambda);\\\tilde\phi&=r\int_{R_0/r}^1g(lr)\sqrt{1-\kappa^2(lr,\theta^2)}\;\d
       l.\nonumber
      \end{align}
\end{enumerate}
\end{remarks}
\subsection{Constructions in incoming region}
\label{Constructions in incoming region}
We introduce for $R\geq 1$ and $\sigma>0$
 \begin{align*}
 &\Gamma^-_{R,\sigma}(\omega)=\{y\in {\mathbb R}^d\;|\;y\cdot \omega\leq (\sigma-1)|y|,\;|y|\geq R\};\; \omega\in S^{d-1},\\
   &\Gamma^-_{R,\sigma}=\{(y,\omega)\in {\mathbb R}^d\times
   S^{d-1}|\;y \in \Gamma^-_{R,\sigma}(\omega)\}.
 \end{align*}

Mimicking the previous procedure, starting from the mixed problem 
\begin{equation}\label{eq:mixed conditions_negative}
\begin{cases}
\ddot y(t) =-\nabla V(y(t))\\
\lambda=\tfrac {1}{2 }\dot y(t)^2 +V_1(y(t))\\
y(-1)=x\\
\omega=-\lim_{t\to -\infty}\omega(t);\;\omega(t)= y(t)/|y(t)|
\end{cases}\;,
\end{equation} cf. \eqref{eq:mixed conditions},
 we may similarly construct a solution $\phi^-(x,\omega,\lambda)$ to the eikonal
equation in some $\Gamma^-_{R,\sigma}(\omega)$. In fact denoting by
$\phi^+(x,\omega,\lambda)$ the solution from Definition
\ref{defn:phasefunc} this amounts to  taking 
\begin{equation}
  \label{eq:rel_phi+-}
 \phi^-(x,\omega,\lambda)=-\phi^+(x,-\omega,\lambda);\;x\in
 \Gamma^-_{R_0,\sigma_0}(\omega) =\Gamma^+_{R_0,\sigma_0}(-\omega). 
\end{equation}
\subsection{Classification of scattering orbits}
\label{Classification of scattering orbits}
The scattering orbits  may be characterized in terms of the
solutions to \eqref{eq:mixed conditions}  and \eqref{eq:mixed
  conditions_negative} as follows.
\begin{prop} Suppose Conditions \ref{assump:conditions1}--\ref{assump:conditions3}.
  For any scattering orbit $x(t)$ with asymptotic
velocities $\omega^{\pm}$ given by \eqref{eq:asymptotic norm222} and
energy  $\lambda\geq 0$ there exists a
(large) $T_0>0$ such that for all $\pm t\geq T\geq T_0$
\begin{align}
 \label{eq:conj1}x(t)&=y(t\mp T\pm 1; x(\pm T),\omega ^{\pm},\lambda),\\
\dot
x(t)&=\nabla_x \phi^{\pm}(x(t),\omega^{\pm},\lambda).\label{eq:conj2}
\end{align}
\end{prop}
\begin{proof} It suffices to look at the case
$t\to +\infty$. 
The proof of \eqref{eq:conj1} is somewhat similar to the proof of
Lemma \ref{lemma:mixed_3}. We introduce
\begin{equation*}
  \tilde z(t)=x(t-1+T)-y_1(t; x(T),\omega ^+,\lambda).
\end{equation*} It needs to be shown that  $\tilde z(t)=z(t;
x(T),\omega ^+,\lambda);\;t\geq{1}$. 

Clearly $\tilde z(1)=0$. We omit  the notation $\omega ^+$ and
$\lambda$. As in the  proof of
Lemma \ref{lemma:mixed_3} it suffices to show \eqref{eq:tilde zin Z_1}--\eqref{eq:tilde zin Z_222} (with $y_1(\cdot)=y_1(\cdot; x(T))$ used to the right in
\eqref{eq:tilde zin Z_2}). 

By Newton's equation 
\begin{equation*}
  \ddot {\tilde z} =-\int_0 ^1 \nabla^2 V_1(l \tilde z+y_1)\{\tilde
  z\}\d l -\nabla V_2(\tilde z+y_1).
\end{equation*} Writing $q=-\int_0 ^1 \nabla^2 V_1(l \tilde z+y_1)\d
  l$ and $\mathcal R=-\nabla V_2(\tilde z+y_1)$ the form is
\begin{equation*}
  \ddot {\tilde z} =q\tilde
  z +\mathcal R,
\end{equation*} or equivalently
\begin{equation}\label{eq:left1}
  (p_t^2+q)\tilde z=-\mathcal R.
\end{equation} 

By \eqref{eq:alpha} we may estimate $\mathcal R$ as follows in terms of
 any non-negative $\kappa <1+\mu+\epsilon_2$
 \begin{equation}
   \label{eq:Rdelt}
  |\mathcal R (t)| \leq Ct^{-\alpha(1+\mu+\epsilon_2-\kappa)}|x(T)|^{-\kappa}.
 \end{equation}

As for the matrix $q$ we claim that indeed it satisfies the condition \eqref{eq:q(t)} with $\epsilon=\bar
\epsilon_1$ provided $T>0$ is large enough. (Notice that the particular case $l=0$ was used in Section
\ref{Mixed problem in the case V_2 not 0}.) To see  this it suffices to 
show that for any $\delta>0$ there  exists $T>0$  large enough such
that 
\begin{equation}\label{eq:lefhhhh}
  t-1\leq (1+\delta) \tilde t(|l\tilde z(t)+y_1(t)|)
\end{equation} uniformly in $t\geq 1$ and $l\in [0,1]$, cf. Condition
  \ref{assump:conditions2}. Define
  $\theta=\theta(t)\in [0,\pi/2]$ by the relation $\cos \theta=x(t)\cdot
  y_1(t)/(|x(t)|\,|y_1(t)|)$ (abusing here and
  henceforth notation $x(t-1+T)\to x(t)$). We may estimate 
  \begin{equation*}
    |lx(t)+(1-l)y_1(t)|\geq \cos (\theta (t)/2)\min (|x(t)|,|y_1(t)|),
  \end{equation*} and use this bound to the upper limit in the
  integral. Since $\theta(t)\to 0$ for  $T\to \infty$ uniformly in 
  $t\geq 1$ we are left with estimating 
\begin{equation}\label{eq:left2a2}
  t-1\leq (1+\delta) \tilde t ((1-\kappa)|x(t)|)
\end{equation} for a sufficiently small $\kappa >0$. Notice that we
  need this also for the particular choice $x(t)=y_1(t)$.

Now since \eqref{eq:left2a2} is valid for $t=1$ it suffices to show
that the derivative
\begin{equation}\label{eq:left2a21}
  (1+\delta) (1-\kappa)\frac{x(t)}{|x(t)|}\cdot \dot x(t)\big (-2V_1(
  (1-\kappa)|x(t)|)\big )^{-1/2}\geq 1.
\end{equation}
Using  Proposition \ref{thm:clas_vel} and elementary estimates we may
see that uniformly in $t\geq 1$ 
\begin{align*}&\lim _{T\to \infty} \frac{x(t)}{|x(t)|}\cdot \frac{\dot
    x(t)}{|\dot x(t)|}=1,\\
&\lim _{T\to \infty}|\dot x(t)|\big (-2V_1(
  |x(t)|)\big )^{-1/2}\geq 1,
\\
  &\lim _{T\to \infty}\big (-2V_1(
  |x(t)|)\big )^{1/2}\big (-2V_1(
  (1-\kappa)|x(t)|)\big )^{-1/2}=1.
\end{align*}  From this we obtain
  \eqref{eq:left2a21}, and hence  \eqref{eq:left2a2} and the above
  assertion for the matrix~$q$.

If $ \lambda >0$ the condition \eqref{eq:q(t)} holds for the matrix
$q$ for any $\epsilon \in
(0,1)$ (provided $T>0$ is sufficiently large).

Next we claim  that  \eqref{eq:left1} is ``solved'' by 
\begin{equation}\label{eq:left2}
  \tilde z=-(p_t^2+q)^{-1}\mathcal R,
\end{equation} in agreement with the theory of Section
\ref{Time--dependent linear force problem }. To see this we
distinguish between the cases $\lambda=0$ and $\lambda>0$. Suppose
first that $\lambda=0$. The
right hand of \eqref{eq:left2} belongs to  $L^2_{-\tilde s}(1,\infty)$
for some  $\tilde s < 1+\tfrac{\bar
   \epsilon_1}{2}$ due to \eqref{eq:Rdelt} (this is similar to the
 argument following \eqref{eq:u_bound}, in fact it holds with $\tilde s=s$). We claim that also 
 \begin{equation}
   \label{eq:2debd}
   \tilde z\in
 L^2_{-\tilde s }(1,\infty) \text{ for some  }\tilde s < 1+\tfrac{\bar
   \epsilon_1}{2}.
 \end{equation}

To show \eqref{eq:2debd} we shall use \eqref{eq:asymptotic norm2} and
 the fact that 
 \begin{equation}
   \label{eq:1bare}
  \bar \epsilon _1> 1-\alpha(\mu+2\epsilon_2) 
 \end{equation}
as
 follows: Abbreviate $\omega^+=\omega$ and decompose
 \begin{equation}
   \label{eq:1tilz}
   \tilde z= (x -x\cdot \omega \,\omega)-( y_1 -y_1\cdot \omega \,\omega)+\tilde z\cdot
   \omega \,\omega.
 \end{equation}

   The  first two
   terms to the right are of the form $O(t^{\alpha-\alpha\epsilon_2})$, cf. \eqref{eq:asymptotic norm2} and
   \eqref{eq:examlog00}. By \eqref{eq:1bare} the function $t^{\alpha-\alpha\epsilon_2}\in
   L^2_{-\tilde s}$ for some  $\tilde s< 1+\tfrac{\bar
   \epsilon_1}{2}$.

As for the third term to the right in \eqref{eq:1tilz} we write 
\begin{equation*}
  \dot {\tilde z}=|\dot x|{\dot x\over |\dot x|} -|\dot
  y_1|{\dot y_1\over |\dot
  y_1|}=(|\dot x|-|\dot
  y_1|)\omega+O(t^{-\alpha{\mu\over 2}-\alpha\epsilon_2}),
\end{equation*} cf. \eqref{eq:examlog000}.

We estimate
\begin{equation*}
 |\dot x|=\sqrt{-2V_1(|x|)}+ O(t^{-\alpha{\mu\over 2}-\alpha\epsilon_2}).
\end{equation*} Combining this with the equation $|\dot
y_1|=\sqrt{-2V_1(|y_1|)}$   and the estimate
\begin{equation*}
  {lx+(1-l)y_1\over |lx+(1-l)y_1|}=\omega +O(t^{-\alpha\epsilon_2}),
\end{equation*} we conclude that 
\begin{equation*}
|\dot x|-|\dot
  y_1|=\tilde q \tilde z \cdot \omega  +O(t^{-\alpha{\mu\over 2}-\alpha\epsilon_2}),
\end{equation*} where
\begin{equation*}
 \tilde q =\int ^1_0 {-V_1'(l\tilde z +y_1)\over \sqrt {-2V_1(l\tilde
 z +y_1)}}\,\d l. 
\end{equation*}

It follows that 
\begin{equation*}
  {\d \over \d t}\tilde z\cdot \omega= \tilde q \tilde z\cdot \omega+O(t^{-\alpha{\mu\over 2}-\alpha\epsilon_2}),
\end{equation*} which in turn yields
\begin{equation}\label{eq:1tilz88}
 \tilde z\cdot \omega= \int _1^t \e^{\int _s ^t \tilde q\,\d t'}
 O(s^{-\alpha{\mu\over 2}-\alpha\epsilon_2})\,\d s.
\end{equation}

Using Condition
  \ref{assump:conditions3} and  \eqref{eq:lefhhhh} we get 
  \begin{equation}
    \label{eq:tides}
    \tilde q(t)\leq \kappa /(t-1)\text{ for some }\kappa
    < 2^{-1}(1+\bar \epsilon_1),
  \end{equation} uniformly in $t\geq1$.

We insert \eqref{eq:tides} into the right hand side of
\eqref{eq:1tilz88}. Invoking \eqref{eq:1bare} we get $\tilde
z\cdot \omega =O(t^\kappa)$; in particular $\tilde z\cdot \omega \in
 L^2_{-\tilde s }(1,\infty)$ for some  $\tilde s < 1+\tfrac{\bar
   \epsilon_1}{2}$ and \eqref{eq:2debd} is proven.

 Finally by combining Lemma \ref{lemma:reso333},  the fact that $\tilde z(1)=0$, 
 \eqref{eq:left1}, \eqref{eq:2debd}  and the statement following
 \eqref{eq:left2} we conlude that  indeed
 \eqref{eq:left2} holds in the case $\lambda=0$.

The case $\lambda>0$ may be treated similarly although the estimates
are simpler in this case. Now it is enough to verify that both sides
of  \eqref{eq:left2} belong to $L^2_{-\tilde s }$ for some  $\tilde s <
3/2$. The right hand of  \eqref{eq:left2} satisfies this by the
same argument as for $\lambda=0$. As for the left hand of
\eqref{eq:left2} the arguments above lead to 
\begin{equation}
  \label{eq:1tilz88330}
   \tilde z-\tilde z\cdot \omega \,\omega=O(t^{1-\alpha\epsilon_2}),
\end{equation} and 
to the represention 
\begin{equation}\label{eq:1tilz8833}
 \tilde z\cdot \omega= \int _1^t \e^{\int _s ^t \tilde q\,\d t'}
 O(s^{-\alpha\epsilon_2})\,\d s.
\end{equation}  

In combination \eqref{eq:1tilz88330} and \eqref{eq:1tilz8833} lead to
$\tilde z=O(t^{1-\alpha\epsilon_2}).$ Consequently indeed $\tilde z\in
L^2_{-\tilde s }$ for some $\tilde s < 3/2$ in the case $\lambda>0$,
and we may conclude \eqref{eq:left2} as before.

Using \eqref{eq:Rdelt}, \eqref{eq:left2} and the theory of Section
\ref{Time--dependent linear force problem } one may now verify
\eqref{eq:tilde zin Z_1}\nobreakdash--\eqref{eq:tilde zin Z_222} (with
the same $s$) for $T>0$ sufficently big, yielding \eqref{eq:conj1} (by
the arguments of the proof of Lemma \ref{lemma:mixed_3}). The
arguments are similar to the proof of Proposition \ref{lemma:mixed_2}.
(Notice for \eqref{eq:tilde zin Z_2} that the estimate
\eqref{eq:alpha} with $x(t)\to y_1(t ; x(T))$ holds uniformly in all
large $T>0$.)

 Clearly \eqref{eq:conj2} follows from \eqref{eq:conj1} and Lemma
 \ref{lemma:mixed_3}.
\end{proof}

\end{document}